\documentclass[10pt]{iopart}

\usepackage{pictex,epsfig, psfrag,rotating} 
\usepackage{amsmath}
\usepackage{amsfonts}

\def\PRL{{\em Phys. Rev. Lett.} }
\def\PRA{{\em Phys. Rev.} A }
\def\PRB{{\em Phys. Rev.} B }

\def\PRE{{\em Phys. Rev.} E }
\def\PR{{\em Phys. Rev.} }
\def\ZPB{{\em Z. Phys.} B }

\def\EPL{{\em Europhys. Lett.} }
\def\JPA{{\em J. Phys.} A }
\def\JPC{{\em J. Phys.} C }
\def\JSP{{\em J. Stat. Phys.} }
\def\PA{{\em Physica} A }
\def\PD{{\em Physica} D }
\def\PTP{{\em Prog. Theor. Phys.} }
\def\ACP{{\em Adv. Chem. Phys.} }

\newcommand{\sst}{\scriptscriptstyle}

\newcommand{\nts}[1]{\tmspace{-}{#1\thinmuskip}{#1\txtmu}}
\newcommand{\notice}[1]{}
\newcommand{\arsinh}{\mathrm{arsinh}}
\newcommand{\arcosh}{\mathrm{arcosh}}

\newcommand{\N}{\mathbb{N}}
\newcommand{\R}{\mathbb{R}}

\newcommand{\fmsgn}{f^{-1}_{\{\sigma\}_n}}
\newcommand{\fsgn}{f_{\{\sigma\}_n}}
\newcommand{\xsgn}{x^*_{\{\sigma\}_n}}
\newcommand{\asgn}{\alpha_{\{\sigma\}_n}}

\newcommand{\supp}{\mathrm{supp} \;}

\begin{document}

\title{Orbits and phase transitions in the multifractal spectrum} 
\author{Thomas Nowotny, Heiko Patzlaff and Ulrich Behn}

\address{Institut f\"ur Theoretische Physik, Universit\"at Leipzig \\ 
Augustusplatz 10, 04109 Leipzig, Germany}

\begin{abstract} 
  We consider the one dimensional classical Ising model in a symmetric
  dichotomous random field. The problem is reduced to a random iterated
  function system for an effective field. The $D_q$-spectrum of the invariant
  measure of this effective field exhibits a sharp drop of all $D_q$ with $q
  < 0$ at some critical strength of the random field. We introduce the
  concept of orbits which naturally group the points of the support of the
  invariant measure. We then show that the pointwise dimension at all points
  of an orbit has the same value and calculate it for a class of periodic
  orbits and their so-called offshoots as well as for generic orbits in the
  non-overlapping case.  The sharp drop in the $D_q$-spectrum is analytically
  explained by a drastic change of the scaling properties of the measure near
  the points of a certain periodic orbit at a critical strength of the random
  field which is explicitly given. A similar drastic change near the points
  of a special family of periodic orbits explains a second, hitherto unnoticed
  transition in the $D_q$-spectrum. As it turns out, a decisive role in this
  mechanism is played by a specific offshoot. We furthermore give rigorous
  upper and/or lower bounds on all $D_q$ in a wide parameter range. In most
  cases the numerically obtained $D_q$ coincide with either the upper or the
  lower bound.  The results in this paper are relevant for the understanding
  of random iterated function systems in the case of moderate overlap in
  which periodic orbits with weak singularity can play a decisive role.
\end{abstract}

\pacs{05.45.Df, 05.50.+q, 75.10.Nr, 05.70.Fh }

\section{Introduction} \label{sec1}
The properties of multifractal measures have attracted a lot of interest over
the past two decades. Multifractals naturally appear in a variety of physical
and mathematical contexts. From the beginning the one dimensional random
field \cite{bruinsma}-\cite{nieuwenhuizenluck} and random exchange
\cite{derrida}-\cite{tanaka2} Ising models were prominent examples. In
treating these systems some reduction scheme for the partition function like
the transfer matrix method \cite{bruinsma, normand, nieuwenhuizenluck,
  derrida} or a method introduced by Ruj\'an \cite{rujan} usually is used
\cite{gyoergyi1}-\cite{pbl} to obtain a random iterated function system (RIFS)
for a local effective field. This leads via a Frobenius-Perron or
Chapman-Kolmogorov equation to an invariant measure which typically is a
multifractal. Similar structures arise in other one dimensional disordered
systems like phonons \cite{schmidt} or electrons \cite{halperin, barnes} in
random potentials, cf.\ also \cite{luckbuch}.

In the early investigations of the random field Ising model the language of
multifractals had not yet been developed and the results focused apart from
calculating the free energy on the structure of the support of the invariant
measure \cite{bruinsma}-\cite{behn3}
and the ground state properties of the system \cite{behnx, behn2, behn3, behn5,
  derrida}. More recently the uniqueness of Gibbs measures and exact ground
state properties were investigated \cite{brzx}. A connection to domain theory
was established in \cite{edalat}.

The invariant measure of
the local effective field in the random field Ising model is a
multifractal \cite{behn3}. In various works the generalized box dimensions
(generalized R\'enyi dimensions) $D_q$ \cite{halsey} of this
measure were calculated for special $q$ \cite{gyoergyi1, evangelou},
with perturbation expansions \cite{behn1, evangelou, bene1, bene2} or
numerical approximations \cite{evangelou, behn6, bene2}. Other authors
focused on different concepts like the order-$q$ free energy and its
fluctuations \cite{tanaka1, tanaka2} or correlation functions
\cite{nieuwenhuizenluck}.

The systematic numerical investigation of the dependence of $D_q$ on the
strength of the local random fields \cite{behn6} revealed the surprising
feature of discontinuities (phase transitions) in the $D_q$ with negative
$q$.

Almost all features of the $D_q$-spectrum have by now been understood
analytically.  The most drastic transition in the $D_q$-spectrum, the sharp
drop of all generalized dimensions $D_q$ with $q < 0$ at some critical field
strength $h_c^{(2)}$, also present in the context of a special model of
neural networks \cite{vanHemmen, behn7}, was explained on a phenomenological
level by the disappearance of deep cuts in the measure density at $h \leq
h_c^{(2)}$ \cite{bvhklz, behn7}.  This disappearance of deep cuts in the
measure density can be explained analytically by close investigation of the
obtained nonlinear RIFS.

In this paper we complete the analysis of the transition begun in \cite{pbl,
  heiko} and explicitly calculate the critical field strength $h_c^{(2)}$ of
the transition.  The result is obtained by generalizing the analysis of the
singularity (pointwise dimension) at fixed points to the singularity of {\em
  orbits}. Further application to a special family of periodic orbits
explains a so far unnoticed smaller drop in the $D_q$-spectrum at a critical
field strength $h_c^{(2a)}$ which became observable because of increased
precision in the numerical generation of the $D_q$-spectrum. Furthermore, the
concept of orbits and their singularity also allows to give bounds on $D_q$
for any $q$ and the exact value of $D_{\pm \infty}$ in a wide parameter
region of the random field strength $h$. The computation of $D_{\pm \infty}$
generalizes earlier results in \cite{evangelou}.

Similar approaches and arguments may be found in the mathematical
literature. In \cite{solomyak} parabolic function
systems with overlaps are considered, \cite{ledrappier} concentrates on
measures obtained by infinite Bernoulli convolutions and \cite{falconerx}
investigates generalized dimensions $D_q$ of measures on general self-affine
sets.

\begin{figure}[t]
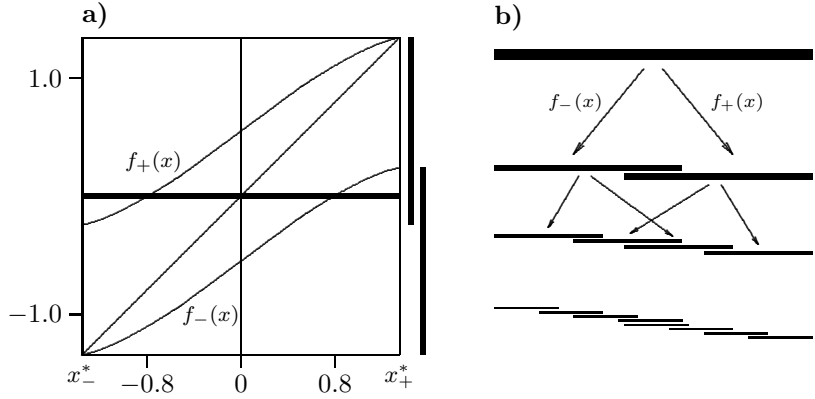

\begin{center}
\mbox{
\beginpicture
\small
\setcoordinatesystem units <44.47pt,44.74pt> point at 1.349 1.341
\setplotarea x from -1.349 to 1.349, y from -1.341 to 1.341
\put {$x^*_{-}$} [t] at -1.349 -1.425
\put {$x^*_{+}$} [t] at +1.349 -1.425
\put {\normalsize \bf a)} [l] at -1.349 1.53
\footnotesize
\put {$f_{+}(x)$} [rb] at -0.5  0.2
\put {$f_{-}(x)$} [lt] at -0.5 -0.92
\normalsize
\axis bottom ticks numbered at  -0.8 0  0.8  / /
\axis left   ticks numbered at  -1.0    1.0  / /
\axis top /
\axis right /
\setquadratic
\plot
-1.349  -0.241
-1      -0.1125
-0.5    0.182
0       0.55
0.5     0.9177
1.0     1.2125
1.349   1.341 /
\plot
-1.349  -1.341
-1.0    -1.2125
-0.5    -0.918
0.0     -0.55
0.5     -0.182
1.0     0.1125
1.349   0.241 /
\setlinear
\plot -1.349 -1.341 1.349 1.341 /
\putrule from 0 -1.341 to 0 1.341
\linethickness=2pt
\putrule from -1.349  0      to 1.349  0
\putrule from  1.45  -0.241  to 1.45  1.341
\putrule from  1.55  -1.341  to 1.55  0.241
\setcoordinatesystem units <120pt,107.1pt> point at -0.3 1.06
\setplotarea x from 0 to 1, y from -0.12 to 1
\put {\bf b)} [l] at 0 1.134
\linethickness=3.5pt
\putrule from .000  1   to 1.000  1
\linethickness=2pt
\putrule from .000 .6    to  .59  .6
\putrule from .41  .57   to 1.00  .57
\linethickness=1pt
\putrule from .000 .36   to  .34   .36
\putrule from .25  .34   to  .59   .34
\putrule from .41  .32   to  .75   .32
\putrule from .66  .30   to 1.00   .30
\linethickness=0.5pt
\putrule from .00  .105  to  .20   .105
\putrule from .14  .09   to  .34   .09
\putrule from .25  .075  to  .45   .075
\putrule from .39  .06   to  .59   .06
\putrule from .41  .045  to  .61   .045
\putrule from .55  .03   to  .75   .03
\putrule from .66  .015  to  .86   .015
\putrule from .80  .0    to 1.00   .0
\arrow <5pt> [.2,.4] from 0.47 0.95 to 0.25 0.65
\arrow <5pt> [.2,.4] from 0.53 0.95 to 0.75 0.65
\arrow <3pt> [.2,.4] from 0.265 0.57 to 0.17 0.39
\arrow <3pt> [.2,.4] from 0.305 0.57 to 0.56 0.36
\arrow <3pt> [.2,.4] from 0.675 0.54 to 0.43 0.37
\arrow <3pt> [.2,.4] from 0.715 0.54 to 0.83 0.33
\scriptsize
\put {$f_{-}(x)$} [rb] at 0.34  0.8
\put {$f_{+}(x)$} [lb] at 0.68  0.8
\normalsize
\endpicture
}
\end{center}
\caption{(a) Mapping in the case of overlapping bands $I_+=f_+(I)$ and
  $I_-=f_-(I)$. In (b) the first few images of the interval $I$ are
  shown. The increasing complexity of the band structure is
  obvious ($h=0.55$, $\beta=1$, $J=1$).}
\label{fig1}
\end{figure}

In the following we consider the one-dimensional random field Ising model
\cite{bruinsma}-\cite{nieuwenhuizenluck} with the Hamiltonian
\begin{equation}
H_N=- J \sum_{i=1}^{N-1} s_is_{i+1} - \sum_{i=1}^{N} h_is_i,
\end{equation}
in which $s_i$ denotes the classical spin at site $i$ which takes values
$1$ or $-1$ and $J$ is the exchange energy of adjacent spins. The local
magnetic fields $\{h_i\}$ at the sites $i= 1, \dots, N$ are
independent identically distributed random variables. We restrict
ourselves to dichotomous symmetric distributions, i.e. to probability
densities with Dirac masses at $\pm h$,
\begin{equation}
\rho (h_i)= \frac{1}{2} \delta (h_i-h)+\frac{1}{2} \delta (h_i+h), \quad h
\in \R^+ . \label{hndist}
\end{equation}
An iterative reformulation of the canonical partition function yields the
partition function of a single spin in an effective external random field
$x_1^{(N)}$ (the effective field at site $1$ in a chain of $N$ spins) which
is given by an iterative map \cite{gyoergyi1},
\begin{align}
  x_i^{(N)} &= h_i + A(x_{i+1}^{(N)}) , \quad x^{(N)}_{N+1}= 0 \label{mapping} \\
  A(x) &= \frac
  1{2\beta }\ln \Big(\frac{\cosh \beta (x+J)}{\cosh \beta (x-J)}\Big),
  \nonumber
\end{align}
where $\beta$ denotes the inverse temperature.\footnote{Note, that we find it
  convenient to use a slightly different notation than in previous work as
  e.g. \cite{behn2, behn3, pbl}.} The iteration is illustrated in figure
\ref{fig1}.  As a shorthand we introduce
\begin{align}
  f_\sigma(x):= \sigma h +A(x), \quad \sigma \in \{+,-\}, \label{fsigma}
\end{align}
such that the recursion (\ref{mapping}) reads $x_i^{(N)}=
f_{\sigma_i}(x_{i+1}^{(N)})$ with $h_i =: \sigma_i h$.  By the reformulation
of the canonical partition function we are thus led to a RIFS with smooth,
strictly monotonously growing, contractive functions $\{f_+,f_-\}$= $\{A+h,
A-h\}$ and probabilities $\{p_+, p_-\}= \{\frac{1}{2},\frac{1}{2}\}$.  When
viewing (\ref{mapping}) as a RIFS we will also write $x_n$ instead of
$x_i^{(N)}$ for the value of the effective field after $n=N-i+1$ iterations.
Please note that the transition from $N$ to $N+1$ spins implies prepending
functions to the composition of functions, i.e.  we need to consider $x_n=
f_{\sigma_1} \circ \ldots \circ f_{\sigma_n} (x)$, $x_{n+1}= f_{\sigma_1}
\circ \ldots \circ f_{\sigma_n} \circ f_{\sigma_{n+1}} (x)$, etc.

We introduce a symbolic dynamic in the obvious way: Let $\Sigma_n$ be the set
of finite sequences $\{\sigma\}_n$ of $n$ symbols $\sigma_i \in \{+,-\}$,
$i=1, \ldots, n$  and
$\Sigma_\infty$ the set of all infinite sequences
$\{\sigma\}$. Given $\{\sigma\}$ we will write $\{\sigma\}_n$ for the {\em
  head} of the $n$ leftmost symbols in $\{\sigma\}$. By $\fsgn$ we denote the
composition of the $n$ functions $f_{\sigma_i}$, $i= 1, \ldots, n$, i.e.
$\fsgn= f_{\sigma_1} \circ f_{\sigma_2} \circ \ldots \circ f_{\sigma_n}$. The
above mentioned properties of $f_+$ and $f_-$ imply the following facts:
\begin{itemize}
  \item The fixed points $x^*_+$ and $x^*_-$ with $f_+(x^*_+)= x^*_+$ and
    $f_-(x^*_-) = x^*_-$ exist.
  \item The interval $I= [x^*_-, x^*_+]$ is the smallest interval with
    $f_\sigma (I) \subseteq I$.
  \item The limit $\lim_{n \to \infty} \fsgn(x_0)$ exists for any
    $\{\sigma\} \in \Sigma_\infty$ and $x_0 \in I$ and does not depend on
    $x_0$. We thus can define $x^*_{\{\sigma\}} := \lim_{n \to \infty}
    \fsgn(x_0)$ and the (constant) function $f_{\{\sigma\}}: I \to I$,
    $f_{\{\sigma\}} (x) := x^*_{\{\sigma\}}$. Through this definition
    $x^*_{\{\sigma\}}$ is the unique fixed point of $f_{\{\sigma\}}$.
  \item We denote the fixed points of finite compositions $f_{\{\sigma\}_n}$
    by $\xsgn$. If $\{\sigma\}$ is periodic with period $n$ then
    $x^*_{\{\sigma\}}= \xsgn$.
  \item We name the $n$-fold images $I_{\{\sigma\}_n}:=f_{\{\sigma\}_n} (I)$
    of the invariant interval $I$
    {\em bands} (of order $n$). For given $\{\sigma\}$ they have the
    inclusion property $I_{\{\sigma\}_n} \subset I_{\{\sigma\}_m}$ for $n >
    m$. Furthermore, the construction implies $x^*_{\{\sigma\}} \in
    I_{\{\sigma\}_n}$ for all $n$.
  \item If the first order bands $I_+$ and $I_-$ do not overlap, $I_+ \cap
    I_- = \emptyset$, none of the higher order bands overlap,
    $I_{\{\sigma\}_n} \cap I_{\{\tilde{\sigma}\}_n} = \emptyset$. We call this
    the non-overlapping case and denote the first order gap by $\Delta:=
    [f_-(x^*_+), f_+(x^*_-)]$. 
    The inverse statement is also true. If the first order bands overlap the
    corresponding higher order bands also overlap. This situation is
    illustrated in figure \ref{fig1}. We denote the first order overlap by
    $O := I_+ \cap I_-$.
\end{itemize}
%
The RIFS (\ref{hndist},\ref{mapping}) induces a
probability density $p_n$ for the effective field $x_n$. We write $P_n(x) :=
\int_0^x p_n(\xi) d\xi$ for the corresponding distribution function and $\mu_n$
for the corresponding measure. $P_n$, $p_n$ and $\mu_n$ can iteratively be
constructed using the Frobenius-Perron (Chapman-Kolmogorov) equation induced
by (\ref{mapping}) which
reads for $P_n$
\begin{align}
  P_{n}(x) = \int \! dh \; \rho (h)P_{n-1}\big(A^{-1}(x-h)\big) =
  \sum_{\sigma=\pm} \frac{1}{2} P_{n-1}\big(f_{\sigma}^{-1}(x)\big)
  \label{frobenius}
\end{align}
with $P_0(x)=\Theta (x)$, $\Theta$ being the Heaviside function. This choice
of $P_0$ encodes the free boundary conditions chosen in (\ref{mapping}).
Again the properties of $f_+$ and $f_-$ imply some direct consequences:
\begin{itemize}
\item The Frobenius-Perron equation has a unique fixed point $P_\infty$
  ($\mu_\infty$) \cite{behnx, hutchinson} and the invariant measure
  $\mu_\infty$ is ergodic.
\item The reiterated application of the Frobenius-Perron equation to an {\em
    arbitrary} initial measure $\mu_0$ (distribution $P_\infty$) converges to
  $\mu_\infty$ ($P_0$) in Hutchinson topology or, as $I$ is compact,
  equivalently in the weak topology of measures \cite{hutchinson}. Thus,
  $\mu_\infty$ ($P_\infty$) is the measure (distribution) of the effective
  field $x$ in the thermodynamic limit $N \to \infty$ for {\em any} boundary
  condition.
\item The explicit form of the $n$-th iterate $P_n(x)$ reads
  \begin{align}
    \label{density}
    P_n(x) &= \sum\limits_{\{\sigma \} _n} \frac{1}{2^n}\,P_0\big(f_{\{\sigma
      \}_n}^{-1}(x)\big) ,
  \end{align}
  which is in the limit $n \to \infty$ a path integral in the space of
  symbolic dynamics. In the non-overlapping case the sum on the right hand
  side of (\ref{density}) has only one term for each $x$. In the
  overlapping case typically more than one term contributes for each $x$ .
  There are however
  $x \in \supp \mu_\infty$ for which still only one term contributes. 
\item The support $\supp \mu_\infty \subseteq I$ is the attractor of the
  RIFS $\{f_+, f_-\}$ \cite{hutchinson}.
\item For any $x \in \supp \mu_\infty$ one can find a sequence $\{\sigma\}$
  with $x= x^*_{\{\sigma\}}$. In the non-overlapping case this relation
  between $\Sigma_\infty$ and $\supp \mu_\infty$ is one to one.
\end{itemize}
In the following we closely investigate the multifractal
properties of the invariant measure $\mu_\infty$. To this end we study the
generalized box counting dimensions
\begin{align}
  D_q= \frac{1}{q-1} \lim_{\varepsilon \to 0} \frac{\ln(\sum_{i} \mu_i^q)}{\ln
    \varepsilon} , \label{efgh}
\end{align}
where $\mu_i= \mu_\infty(B_\varepsilon(x_i))$ are the measures of boxes
(intervals) of
size $\varepsilon$ covering $\supp \mu_\infty$, and the pointwise dimension
\begin{align}
  D_p(x)= \lim_{\varepsilon \to 0} \frac{\ln \mu_\infty(B_\varepsilon(x))}
  {\ln \varepsilon} \label{ptwisedim}
\end{align}
at individual points $x \in \supp \mu_\infty$. We will synonymously use the
{\em singularity} $\alpha:= D_p -1$. The intricate interplay
between these quantities will explain the transitions in the $D_q$-spectrum
mentioned above. The obtained characterization
of $\mu_\infty$, the measure of the effective field $x$, is a necessary
prerequisite for the more complicated treatment of the distribution of
physical quantities like the local magnetisation which will follow in a later
publication.

The paper is organised as follows. In subsection \ref{knownsubsec} we briefly
review the known results about prominent features of the $D_q$-spectrum. In
section \ref{orbitsec} the concept of orbits and their singularity is
introduced and in \ref{singorbitsec} this singularity is calculated for a
class of periodic orbits and their so-called offshoots. In subsection
\ref{arbisec} we calculate the generic singularity of arbitrary orbits in the
non-overlapping case. In subsection \ref{hc2sec} we treat the overlapping
case and discuss effects occurring if points of an orbit enter the overlap.
The results are used to explain the transition in the $D_q$-spectrum at
$h_c^{(2)}$ and calculate $h_c^{(2)}$ as a function of temperature $T$ and
coupling strength $J$ explicitly. We then apply a similar analysis to the
transition at $h_c^{(2a)}$ in subsection \ref{hc2asec}. In section
\ref{boundsec} we give the extended lower and upper bounds on the
$D_q$-spectrum obtained from the analysis of the singularity of specific
orbits.  Finally, some conclusions are drawn in section \ref{concludesec}.

\subsection{Known results} \label{knownsubsec}
In this subsection we summarize in short previous work on phase transitions
in the invariant measure $\mu_\infty$.  For large $h$ the support of
$\mu_\infty$ is non-connected and similar to a multi-scale Cantor set
\cite{behn3}.

At a critical value $h_c^{(1)}$ of $h$ the support of
$\mu_\infty$ becomes connected for all $h \leq h_c^{(1)}$ \cite{normand}. The
value of $h_c^{(1)}$ is determined by the overlap condition for the first
bands, $f_-(x^*_+)= f_+(x^*_-)$. This results in \cite{behn6} \begin{figure}[t]
\begin{center}
  \psfrag{D}{$D_q$}
  \psfrag{0}{0}
  \psfrag{0.2}{0.2}
  \psfrag{0.4}{0.4}
  \psfrag{0.5}{0.5}
  \psfrag{0.6}{0.6}
  \psfrag{0.8}{0.8}
  \psfrag{1}{1}
  \psfrag{1.5}{1.5}
  \psfrag{2}{2}
  \psfrag{h}{$h$}
  \psfrag{q=m}{\parbox{1cm}{\flushright \tiny$q=-\infty$}}
  \psfrag{q=-20}{\parbox{1cm}{\flushright \tiny$q=-20$}}
  \psfrag{q=-6}{\parbox{1cm}{\flushright \tiny$q=-6$}}
  \psfrag{q=-3}{\parbox{1cm}{\flushright \tiny$q=-3$}}
  \psfrag{q=-2}{\parbox{1cm}{\flushright \tiny$q=-2$}}
  \psfrag{q=-1}{\tiny$q=-1$}
  \psfrag{q=0}{\tiny$q=0$}
  \psfrag{q=1}{\tiny$q=1$}
  \psfrag{q=2}{\tiny$q=2$}
  \psfrag{q=4}{\tiny$q=4$}
  \psfrag{q=20}{\tiny$q=20$}
  \psfrag{q=p}{\tiny$q=\infty$}
  \psfrag{hc1}{\small$\!\!\!\!\!\begin{array}{l} h_c^{(4)} \\ \downarrow
    \end{array}$}
  \psfrag{hc2}{\small$\!\!\!\!\!\begin{array}{l} h_c^{(3)} \\ \downarrow
    \end{array}$}
  \psfrag{hc3}{\small$\!\!\!\!\!\begin{array}{l} h_c^{(2)} \\ \downarrow
    \end{array}$}
  \psfrag{hc3a}{\small$\!\!\!\!\!\begin{array}{l} h_c^{(2a)} \\ \downarrow
    \end{array}$}
  \psfrag{hc4}{\small$\!\!\!\!\!\begin{array}{l} h_c^{(1)} \\ \downarrow
    \end{array}$}
 \epsfig{file=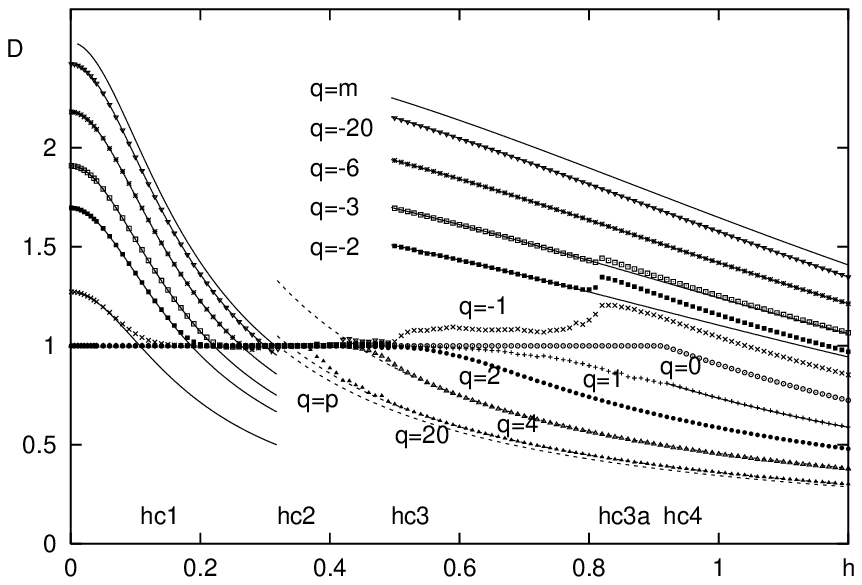, width=\textwidth}
\end{center}  
\caption{
  Generalized fractal dimensions $D_q$ of the invariant measure of the local
  effective field with $q=-20$, $-6$, $-3$, $-2$, $-1$, $0$, $1$, $2$, $4$,
  $20$ versus the amplitude of the local random field $h$ ($\beta=1$, $J=1$).
  The results were computed in the thermodynamic formalism using the new
  natural partition introduced by Behn and Lange \cite{behn6}. The
  significance of the critical values $h_c^{(n)}$, $n=1,\dots ,4$, and
  $h_c^{(2a)}$ is explained in the text. The solid and dashed lines are exact
  lower and upper bounds on $D_q$ respectively which are obtained in section
  \ref{boundsec} except for the solid line coinciding with the values of
  $D_1$ for $h > h_c^{(1)}$ which was obtained from
  (\ref{dpformula}).}
\label{fig2}
\end{figure}   
%

\begin{align}
h_c^{(1)} = \frac{1}{2\beta} \arcosh\big((e^{2\beta J}-1)/2\big) .
\end{align}
The transition can be seen in the densities $p_n$ of the approximations
$\mu_n$ of the invariant measure $\mu_\infty$ \cite{behn6}.
In the
$D_q$-spectrum the transition is visible as the point where $D_0$ becomes $1$,
cf. figure \ref{fig2}.

At $h_c^{(3)} \leq h_c^{(1)}$ the invariant measure density jumps from
infinity to zero at the boundary $x^*_+$ and $x^*_-$ of $I$ \cite{behn6}. This
is an effect solely depending on the scaling of the measure at the fixed
points $x^*_-$ and $x^*_+$.
The measure density $p_n(x^*_\pm)$ diverges for
$f'_\pm(x^*_\pm) < \frac{1}{2}$ and converges to zero for $f'_\pm(x^*_\pm) >
\frac{1}{2}$ leading to the critical value \cite{behn6}
\begin{align}
h_c^{(3)} = \frac{1}{\beta} \arsinh\big(2^{-\frac{3}{2}} (1-9e^{-4\beta
J})^{\frac{1}{2}}\big) .
\end{align}
The transition is again visible in numerically generated $p_n$
\cite{behn6} as
well as in the $D_q$-spectrum as the value of $h$
for which $D_{-\infty}$ begins to grow again for decreasing $h$,
cf. figure \ref{fig2}. As was shown in \cite{behn7}, the generalized
fractal dimension $D_{-\infty}$ has the value $D_{-\infty}= 1$ at this
point and $D_{-\infty} > 1$ for $h < h_c^{(3)}$.

The last of the transitions which are already well understood occurs at $h_c^{(4)} \leq h_c^{(3)}$ when the
slope of the coarse grained invariant measure density at $x^*_\pm$
jumps from $\mp \infty$ to $0$.
The condition for this is $f'_\sigma (x^*_\sigma) =
2^{-1/2}$ \cite{bvhklz}
resulting in \cite{behn6}
\begin{align}
  h_c^{(4)}= \frac{1}{\beta} \arsinh \big(3 \cdot 2^{-\frac{5}{2}} -
{\textstyle \frac{1}{2}} - (3 \cdot 2^{-\frac{5}{2}} + {\textstyle
\frac{1}{2}}) e^{-4\beta J} \big)^{\frac{1}{2}} .
\end{align}
The transition is 
visible in numerically generated densities $p_n$ \cite{behn6} 
but not in the $D_q$-spectrum, cf. figure \ref{fig2}. Again,
$D_{-\infty}$ can be calculated analytically and takes the value
$D_{-\infty}= 2$. In fact, $D_{-\infty}$ can be
calculated analytically for all $h < h_c^{(3)}$ by considering that the
scaling at the boundary in this case is weaker than at any other point
\cite{evangelou}. This
gives lower and upper bounds on $D_q$ which for $q \to -\infty$ converge
against each other, see section \ref{boundsec}.

Please note that the effects summarized in this subsection only depend on the
measure at the boundary of its support and therefore are not strictly
multifractal effects.

\section{Orbits and their contribution to the invariant
  measure}\label{orbitsec} The orbit to a given symbolic sequence
$\{\sigma\}$ consists of all preimages $\fmsgn (x^*_{\{\sigma\}})$, $n \in
\N_0$. In the case of a periodic sequence $\{\sigma\}=(\{\sigma\}_n)_\infty$
with finite period $n$, the orbit consists of the fixed points of the $n$
functions $f_{\pi\{\sigma \}_n}$ in which $\pi$ denotes a cyclic permutation.
To denote periodic orbits we will write for simplicity $\{\sigma\}_n$ instead
of $(\{\sigma\}_n)_\infty$
The fixed points $\xsgn$ of the periodic orbits are dense in the support of
$\mu_\infty$ \cite{hutchinson}. Furthermore, any point of the support of
$\mu_\infty$ is contained in at least countably infinitely many orbits.

\subsection{Singularity of periodic orbits}
\label{singorbitsec}
For the case of the fixed points $x^*_\pm$ of $f_\pm$ it has been shown
before that their singularity can be calculated explicitly \cite{bvhklz,pbl}.
Fixed points are $1$ - orbits and we generalize this concept to periodic
orbits of arbitrary period length.

\begin{figure}
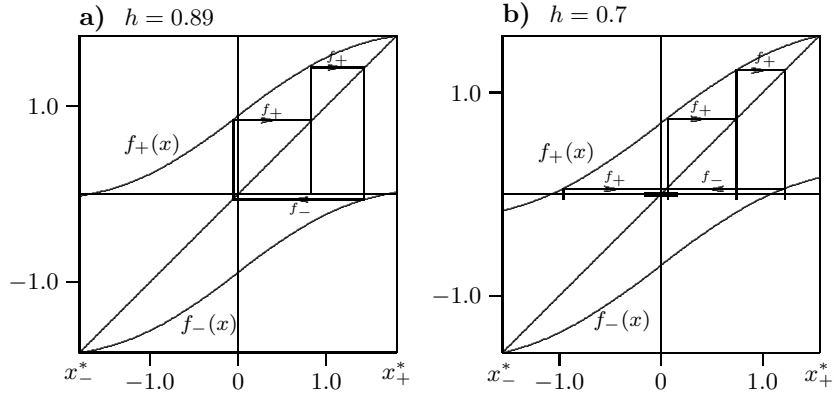

\begin{center}
\mbox{

\beginpicture
\small
\setcoordinatesystem units <33.33pt,33.33pt> point at 2.4 1.8
\setplotarea x from -1.8 to 1.8, y from -1.8 to 1.8
\put {$x^*_{-}$} [t] at -1.8 -1.9
\put {$x^*_{+}$} [t] at +1.8 -1.9
\put {{\normalsize \bf a) } $h=0.89$} [l] at -1.8 2.0
\axis bottom ticks numbered at  -1.0 0 1.0  / /
\axis left   ticks numbered at  -1.0   1.0  / /
\axis top /
\axis right /
\setquadratic
\plot -1.8 -0.02 -1.35 0.087 -0.9 0.278 -0.45 0.557 0.0 0.89 0.45 1.223
0.9 1.502 1.35 1.693 1.8 1.8 /  
\plot -1.8 -1.8 -1.35 -1.693 -0.9 -1.502 -0.45 -1.223 0.0 -0.89 0.45 -0.557
0.9 -0.278 1.35 -0.087 1.8 0.02 /
\setlinear
\plot -1.8 -1.8  1.8 1.8 /
\putrule from 0 -1.8 to 0 1.8
\putrule from -1.8 0 to 1.8 0
\putrectangle corners at -0.0521 0.84   and 0.827 0.0
\putrectangle corners at  0.827  1.44   and 1.43  0.0
\putrectangle corners at  1.43  -0.06   and -0.0521 0.0
\arrow <5pt> [.2,.4] from 0.34 0.84 to 0.41 0.84
\arrow <5pt> [.2,.4] from 1.08 1.44 to 1.17 1.44
\arrow <5pt> [.2,.4] from 0.73 -0.06 to 0.65 -0.06
\tiny
\put {$f_{+}$} [b] at  0.39  0.85
\put {$f_{+}$} [b] at  1.13  1.45
\put {$f_{-}$} [t] at  0.71  -0.1
\footnotesize
\put {$f_{+}(x)$} [rb] at -0.65  0.45
\put {$f_{-}(x)$} [rb] at  0.0  -1.6

\small
\setcoordinatesystem units <38.46pt,38.46pt> point at -2.08 1.56
\setplotarea x from -1.56 to 1.56, y from -1.56 to 1.56
\put {$x^*_{-}$} [t] at -1.56 -1.66
\put {$x^*_{+}$} [t] at +1.56 -1.66
\put {{\normalsize \bf b) } $h=0.7$} [l] at -1.56 1.76
\axis bottom ticks numbered at  -1.0 0 1.0  / /
\axis left   ticks numbered at  -1.0   1.0  / /
\axis top /
\axis top shiftedto y=0
      ticks in length <2pt> unlabeled at -0.96 0.064 0.74 1.22 / /
\axis right /
\setquadratic
\plot -1.56 -0.162 -1.17 -0.038 -0.78  0.155 -0.39 0.409 0 0.7 0.39 0.991
       0.78 1.245 1.17 1.438 1.56 1.562  /      
\plot  -1.56 -1.56 -1.17 -1.438 -0.78 -1.245 -0.39 -0.991 0 -0.7 0.39 -0.409
        0.78 -0.155 1.17 0.038 1.56 0.162 /
\setlinear
\plot -1.56 -1.56  1.56 1.56 /
\putrule from 0 -1.56 to 0 1.56
\putrule from -1.56 0 to 1.56 0
\linethickness=1.5pt
\putrule from -0.162 0 to 0.162 0
\normalgraphs
\putrectangle corners at  0.064  0.74   and 0.74  0.0
\putrectangle corners at  0.74   1.22   and 1.22  0.0
\putrectangle corners at  1.22   0.05   and 0.064 0.0
\putrectangle corners at  0.064  0.05   and -0.96 0.0
\arrow <5pt> [.2,.4] from 0.35 0.74 to 0.42 0.74
\arrow <5pt> [.2,.4] from 0.95 1.22 to 1.02 1.22
\arrow <5pt> [.2,.4] from 0.56 0.05 to 0.49 0.05
\arrow <5pt> [.2,.4] from -0.49 0.05 to -0.43 0.05
\tiny
\put {$f_{+}$} [b] at  0.40  0.77
\put {$f_{+}$} [b] at  1.03  1.25
\put {$f_{-}$} [b] at  0.51  0.08
\put {$f_{+}$} [b] at -0.45  0.08
\footnotesize
\put {$f_{+}(x)$} [rb] at -0.65  0.3
\put {$f_{-}(x)$} [rb] at  -0.1  -1.35
\normalsize
\endpicture
}
\end{center}

\caption{Orbits of period three for different values of $h$. (a) shows
the case where no point of the orbit falls into the overlap region
whereas in (b) the point $x^*_{\{-++\}}$ is in the overlap
region and has therefore two predecessors
($\beta=1$, $J=1$). \label{fig3}}
\end{figure}

 Let $y_i := f^{-1}_{\sigma_i} \circ f^{-1}_{\sigma_{i-1}} \circ
\dots \circ f^{-1}_{\sigma_1} (\xsgn)$, $i=1, \ldots, n$, be the points of the
periodic $n$-orbit defined by $\{\sigma\}_n$. We then have $y_n = y_0 =
x^*_{\{\sigma \}_n}$ because $\xsgn$ is
the fixed point of $f_{\{\sigma \}_n}$ by definition. An example for
a periodic $3$-orbit is shown in figure \ref{fig3}.

In case that no overlap exists or - if it exists - that no point $y_i$ is in the
overlap, the predecessor of each $y_i$ with respect to the iteration of
the Frobenius-Perron equation is uniquely determined to be $y_{i-1}$. There
is therefore only one term in the Frobenius-Perron equation at all $y_i$.  We
now investigate the singularity (pointwise dimension) of $\mu_\infty$ at
$\xsgn$. We assume\footnote{This assumption of strong scaling can be lifted
  in the non-overlapping case in which a generalization of (\ref{alpha}) to
  arbitrary orbits can be proven provided the pointwise dimension of the
  fixed point exists, cf. subsection \ref{arbisec}.}  that the scaling limit
\begin{align}
\lim_{\varepsilon \to 0} \frac{P_\infty(\xsgn +
  \frac{\varepsilon}{2}) -
  P_\infty(\xsgn - \frac{\varepsilon}{2})}{\varepsilon^{\asgn+1}} =: k \label{scale}
\end{align}
exists for some finite $k \neq 0$ and some $\asgn \in \R$. As is
shown in \ref{appa} this implies
\begin{align}
\lim_{\varepsilon \to 0} \frac{P_\infty\big(\fmsgn
  (\xsgn + \frac{\varepsilon}{2})\big) -
  P\big(\fmsgn(\xsgn -
  \frac{\varepsilon}{2})\big)}{\big( ( \fmsgn)' (\xsgn) \varepsilon
  \big)^{\asgn +1}} = k. \label{scale2}
\end{align}
The $n$-fold iteration of the Frobenius-Perron equation yields
\begin{align}
& P_\infty(\xsgn+\textstyle\frac{\varepsilon}{2})-P_\infty(\xsgn-
\textstyle\frac{\varepsilon}{2})
\nonumber \\ 
&= \frac{1}{2^n}
\Big(P_\infty\big(\fmsgn(\xsgn +
\textstyle \frac{\varepsilon}{2})\big)- 
P_\infty\big(\fmsgn(\xsgn -
\textstyle \frac{\varepsilon}{2})\big) \Big) .
\end{align}
We denote the expression on the left hand side by $X$ and the one on
the right hand side by $Y/2^n$ and thus have
$X/Y = 1/2^n$.
Inserting $1 = k/k$ and using (\ref{scale}) and (\ref{scale2}) with the
introduced notation $X$ and $Y$ we get
\begin{align}
  \lim_{\varepsilon \to 0} \bigg( \frac{X}{Y} \cdot \frac{Y\big/\big( (
    \fmsgn)' (\xsgn) \varepsilon
    \big)^{\asgn
      +1}}{X\big/\varepsilon^{\asgn +1}} \bigg) =
  \frac{1}{2^n} .
\end{align}
Most terms immediately cancel and we get
\begin{align}
\big(( \fmsgn)' (\xsgn)\big)^{\asgn +1} = 2^n.
\end{align}
Therefore
\begin{align}
   \asgn + 1 = \frac{n\ln2}
{\ln\big((\fmsgn)'(\xsgn)\big)} 
\end{align}
and with $(\fmsgn)'(\xsgn)=
\big(f'_{\{\sigma\}_n}(\fmsgn (\xsgn))\big)^{-1} =
\big(f'_{\{\sigma\}_n}(\xsgn)\big)^{-1}$ we get
\begin{align}
\asgn= -1 - \frac{n \ln 2}{\ln\big(f'_{\{\sigma\}_n}
  (\xsgn)\big)} .
\end{align}
This equation is invariant under cyclic permutation of $\{\sigma\}_n$ such
that the scaling behaviour of the invariant measure at all $y_i$, $i= 1,
\dots, n$, is given by the same H\"older exponent $\asgn$. In other words,
any point of the orbit has the same pointwise dimension $D_p=
\alpha_{\{\sigma\}_n} -1$. We therefore call $\asgn$ the {\em singularity of
  the orbit} $\{\sigma\}_n$. If $\asgn < 0$ the measure has a positive (i.e.
strong) singularity at all $y_i$ and if $\asgn > 0$ the singularity is
negative (i.e.  weak).

As $f'_{\{\sigma\}_n}(\xsgn) =
\prod_{i=1}^{n} f'_{\sigma_{i}}(y_{i-1})$, the
derivatives $f'_{\sigma_i} (y_{i-1}) = A'(y_{i-1})$ at the points
$y_{i-1}$ of the periodic orbit determine the singularity
$\asgn$ and we finally have
\begin{align}
\asgn= -1 - \frac{n \ln 2}{\sum_{i=1}^n \ln
A' (y_i)} .
\label{alpha}
\end{align}
A short calculation using the Frobenius-Perron equation (\ref{frobenius})
shows that the sin\-gu\-la\-ri\-ties of arbitrary points $x$ and
$f_{\sigma}(x)$ are the same provided $f_\sigma(x)$ is not in the overlap,
cf. \ref{appb}. The argument can be iterated such that the measure has the
same singularity at any $x$ and all its images $f_{\{\tilde{\sigma}\}_m}(x)$
for any $\{\tilde{\sigma}\}$ for which no point
$f_{\{\tilde{\sigma}\}_i}(x)$, $i= 1, \ldots m$ is in $O$. Therefore not only
the singularities at all points of a periodic orbit are the same but also the
singularities at all points of non-periodic orbits in case the orbit also
does not touch the overlap $O$.  For orbits of the form
$\{\tilde{\sigma}\}_m(\{\sigma\}_n)_\infty$ we know that this singularity is
the singularity of $\{\sigma\}_n$, the periodic tail. We call non-periodic
orbits of this type {\em offshoots} of the corresponding periodic orbit. The
roles played by the head and the tail of a symbolic sequence $\{\sigma\}$ are
summarized in table \ref{table1}. Note that the choice of the length of the
head is arbitrary, in a sense. Similar structures have been considered in
\cite{barnes}.\begin{table}
\begin{center}
\renewcommand{\arraystretch}{1.5}
\begin{tabular}{|l l|}
\hline head (finite) & tail (infinite) \\
\hline & \\[-3.85ex] \hline \multicolumn{2}{|l|}{generic} \\
\hline $+ - + + + - - + -$ & $+ - + + + - - + - + + - - - +$ \ldots \\[-0.15cm]
$\{\sigma\}_n$ & $\{\tilde{\sigma}\}$ \\
\hline & \\[-3.85ex] \hline \multicolumn{2}{|l|}{periodic (here period 3)} \\
\hline $- + + - + + - + +$ &  $- + + - + + - + + - + + - + + $
\ldots \\[-0.15cm]
$\{\sigma\}_9 = (- + + )_3$ & $(- + +)_\infty$ \\
\hline & \\[-3.85ex] \hline \multicolumn{2}{|l|}{offshoot (arbitrary head, periodic tail)} \\
\hline $+ - + + - - - + +$ & $- + + - + + - + + - + + - + + $ \ldots \\[-0.15cm]
$\{\sigma\}_n$ & $(- + +)_\infty$ \\
\hline & \\[-3.85ex] \hline \parbox[t]{3.2cm}{head determines the interval
  $I_{\{\sigma\}_n} \ni x^*_{\{\sigma\}}$} \rule[-0.8cm]{0cm}{0.8cm} & tail determines scaling at
  $x^*_{\{\sigma\}} $  \\
\hline
\end{tabular}
\end{center}
\caption{Illustration of the terminology of symbolic sequences and the
  different  roles played by head and tail. \label{table1}}
\end{table}


\subsection{Singularity of generic orbits in the non-overlapping case}
\label{arbisec} 
In the last subsection we have seen that the fact that if the predecessor of
each point of an orbit is unique we can explicitly calculate the
singularity of the orbit as a function of the derivative of $A$ at the points
of the orbit. In the non-overlapping case, $O = \emptyset$, the
uniqueness of predecessors holds for any point in $\supp \mu_\infty$. It is
therefore natural to try to extend (\ref{alpha}) to generic non-periodic orbits.

As a first step it is not difficult to show that if
\begin{align}
  \lim_{n \to \infty} \frac{1}{n} \sum_{i=1}^{n} \ln A' (x_i^{\sst (n)})
  \label{indepsum}
\end{align}
exists for one choice of $x_i^{\sst (n)} \in I_{\{\sigma\}_i}$ for a given
$\{\sigma\}$, then it exists for any such choice and is independent of the
particular choice made, cf. \ref{app2}.

Let $x \in \supp \mu_\infty$ be such that (\ref{indepsum}) exists for the
corresponding symbolic sequence $\{\sigma\}$. To calculate the limit
$\varepsilon \to 0$ in (\ref{ptwisedim}) it is sufficient to consider
$\varepsilon_n:= |I_{\{\sigma\}_n}|$ as
$\frac{\varepsilon_{n+1}}{\varepsilon_n} \geq \min\{A'(x) | x \in I\} =
A'(x^*_+) > 0$ holds. We know from the general properties that $x=
x^*_{\{\sigma\}} \in I_{\{\sigma\}_n}$ for all $n \in \N$. Because of the
choice of $\varepsilon_n$ we have $B_{\varepsilon_n}(x) \supset
I_{\{\sigma\}_n}$ such that $\mu_\infty (B_{\varepsilon_n}) \geq \mu_\infty
(I_{\{\sigma\}_n}) = \frac{1}{2^n}$ leading to
\begin{align}
  \frac{\ln \mu_\infty (B_{\varepsilon_n})}{\ln \varepsilon_n} \leq \frac{-n
    \ln 2}{\ln \varepsilon_n}. \label{dpsub}
\end{align}
On the other hand we also can choose $\varepsilon'_n:= |\fsgn(\Delta)|$. The
interval $I_{\{\sigma\}_n}$ is neighboured by two gaps. One of the
neighbouring gaps is always $f_{\{\sigma\}_{n-1}}(\Delta)$, the other is
either $f_{\{\sigma\}_m}(\Delta)$ with $m < n-1$ or it is the complement of
$I$. By contractivity of the RIFS we have $|f_{\{\sigma\}_m}(\Delta)| >
|\fsgn(\Delta)|$ for $m < n$, such that in either case the smallest gap
neighbouring $I_{\{\sigma\}_n}$ is $f_{\{\sigma\}_{n-1}}(\Delta)$. This
implies $ \mu_\infty(B_{\varepsilon'_n}) \leq \mu_\infty(I_{\{\sigma\}_n}) =
\frac{1}{2^n}$ because $B_{\varepsilon'_n}$ can not bridge any of the
neighbouring gaps and thus only intersects $I_{\{\sigma\}_n}$. Therefore,
\begin{align}
  \frac{\ln \mu_\infty (B_{\varepsilon'_n})}{\ln \varepsilon'_n} \geq \frac{-n
    \ln 2}{\ln \varepsilon'_n}.
  \label{dpsup}
\end{align}
Using the mean value theorem for $\fsgn$ we obtain
\begin{align}
  \varepsilon_n &= |I_{\{\sigma\}_n} | = \fsgn(x^*_+) - \fsgn(x^*_-) \\
  &=
  \big(\fsgn\big)' (x_0^{\sst (n)}) \, (x^*_+ - x^*_-) = \prod_{i=0}^{n-1}
  A'\big(f_{\{\sigma\}_i}(x_0^{\sst(n)})\big)\; |I| \label{eps1}
\end{align}
for some $x_0^{\sst(n)} \in I$.
In the same fashion we get
\begin{align}
  \varepsilon'_n = \prod_{i=0}^{n-1}
  A'\big(f_{\{\sigma\}_i}({x'_0}^{\sst (n)})\big) \; |\Delta| \label{eps2}
\end{align}
for some other ${x'_0}^{\sst (n)} \in I$.  Taking (\ref{dpsub}) and
(\ref{eps1}) we obtain
\begin{align}
  \limsup_{n \to \infty} \frac{\ln \mu_\infty
    (B_{\varepsilon_n})}{\ln \varepsilon_n} &\leq \limsup_{n \to \infty}
  \frac{-\ln 2}{\frac{1}{n} \sum_{i=0}^{n-1} \ln
    A'\big(f_{\{\sigma\}_i}(x^{\sst (n)}_0)\big) + \frac{1}{n} \ln |I|} \\
  &=
  \lim_{n \to \infty} \frac{-n \ln 2}{\sum_{i=1}^n \ln
    A'\big(f_{\{\sigma\}_i}^{-1} (x^*_{\{\sigma\}})\big)} \label{f1}
\end{align}
while using (\ref{dpsup}) and (\ref{eps2}) yields
\begin{align}
  \liminf_{n \to \infty} \frac{\ln \mu_\infty (B_{\varepsilon'_n})}{\ln
    \varepsilon'_n} &\geq \liminf_{n \to \infty} \frac{-\ln 2}{\frac{1}{n}
    \sum_{i=0}^{n-1} \ln A'\big(f_{\{\sigma\}_i}({x'_0}^{\sst (n)})\big) +
    \frac{1}{n} \ln |\Delta|} \\ &= \lim_{n \to \infty} \frac{-n \ln
    2}{\sum_{i=1}^n \ln A'\big(f_{\{\sigma\}_i}^{-1}
    (x^*_{\{\sigma\}})\big)}.
  \label{f2}
\end{align}
In both (\ref{f1}) and (\ref{f2}) the existence of the limit in
(\ref{indepsum}) and the independence of (\ref{indepsum}) of the
points $x_i^{\sst (n)} \in I_{\{\sigma\}_i}$ was used to replace $\liminf$
and $\limsup$ by $\lim$ and to
substitute the points of the orbit $\{\sigma\}$ for $x_0^{\sst (n)}$ and
${x'_0}^{\sst (n)}$ respectively. From (\ref{f1}) and (\ref{f2}) we
immediately get that $D_p$ exists and is given by
\begin{align}
  D_p= \lim_{n\to \infty} \frac{-n \ln 2}{\sum_{i=1}^n \ln
    A'\big(f_{\{\sigma\}_i}^{-1} (x^*_{\{\sigma\}})\big)} . \label{dpformula}
\end{align}
Elton's ergodic theorem \cite{elton} implies that our assumption of the
existence of (\ref{indepsum}) holds for almost all $\{\sigma\}$ which
corresponds to $\mu_\infty$ - almost sure existence of the pointwise dimension
$D_p(x)$. Elton's theorem together with (\ref{dpformula}) further implies
that for $\mu_\infty$ - almost all $x$ the pointwise dimension takes the
common value
\begin{align}
  D_p(x)
  \stackrel{\text{$\mu_\infty$-a.s.\raisebox{-0.15cm}{\rule{0cm}{0.3cm}}}}{=}
  \frac{- \ln 2}{\int \big(\ln A'(\xi)\big) 
    \mu_\infty(d\xi)} =: \overline{D}_p . \label{aaconv}
\end{align}
This result has direct consequences for the information dimension
$D_1$. Proposition 2.1. in \cite{young} implies that
\begin{align}
  D_H\big(\{ x \in I: D_p(x)= \overline{D}_p \}\big) = \overline{D}_p
  \label{hausdorff}
\end{align}
in which $D_H$ denotes the Hausdorff-dimension. General properties of the
multifractal $f(\alpha)$-spectrum imply that $\alpha_1$ is the only fixed
point of $f(\alpha)$ and that $\alpha_1= D_1$, cf. \cite{falconer}. Therefore
(\ref{hausdorff}) implies $D_1= \overline{D}_p$. This fact is illustrated by
the solid line in figure \ref{fig2} coinciding with the numerically obtained
values of $D_1$ for $h > h_c^{(1)}$ which was obtained through calculation
of (\ref{aaconv}) using Edalat's $R$-integration \cite{edalatR}. (It also is
a simple exercise to check on a computer that using the first $10^5$ digits
of the dual representation of $\pi$ or $e$ as a symbolic sequence, the value
of $\alpha$ obtained from (\ref{dpformula}) is also exactly the value $D_1 -
1$ in figure \ref{fig2}.  Of course, the use of a random number generator
instead of $\pi$ or $e$ yields the same result with probability one.)

Please note that the restriction to almost all
$\{\sigma\}$ in the above is necessary as the sum in (\ref{indepsum}) does
not converge for all $\{\sigma\}$. A simple example in which it does not
converge is a sequence of bulks of plus and minus signs of ever increasing
length. The length of the bulks can be chosen such that the sum in
(\ref{indepsum}) keeps oscillating for any size of $n$.

\subsection{Singularity of orbits in the overlapping case}\label{hc2sec}
\begin{figure}
\begin{center}
  \psfrag{xlabel}{\raisebox{-0.2cm}{iteration depth}}
  \psfrag{ylabel}{\turnbox{180}{\raisebox{-0.3cm}{$\nts{20}$
        images or preimages of $x$}}}
  \psfrag{xm}{$\nts{8}$\parbox{1cm}{\flushright \raisebox{0.2cm}{$x_-$}}}
  \psfrag{xp}{$\nts{8}$\parbox{1cm}{\flushright \raisebox{0.2cm}{$x_+$}}}
  \psfrag{overlap}{overlap} \psfrag{s1}{\small $\overline{\alpha}$}
  \psfrag{s2}{\small $\overline{\alpha}$} \psfrag{s3}{\small
    $\nts{10}\alpha_{\{+-\}}$} \psfrag{s4}{\small $\overline{\alpha}$}
  \psfrag{s5}{\small $\overline{\alpha}$}
  \psfrag{x}{{\boldmath$x$}}
  \psfrag{y}{{\boldmath$y$}}
  \epsfig{file=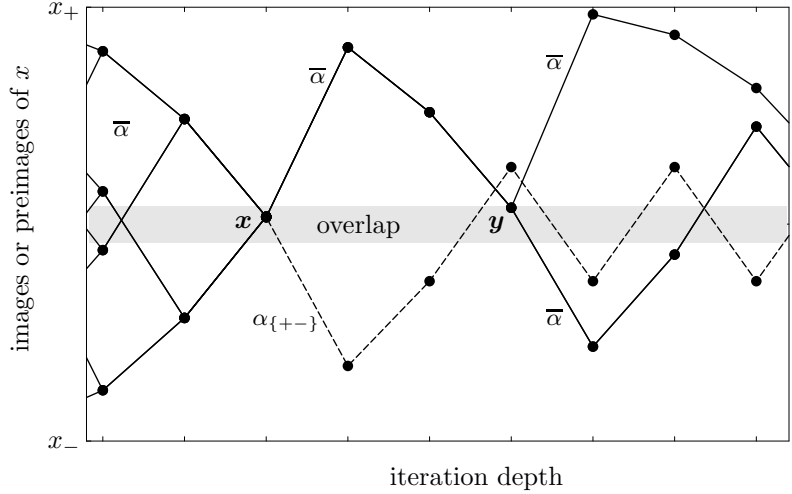, width=0.8\textwidth}
\end{center}
\caption{This figure illustrates the intricate structure of orbits and how
  weak singularities are superseded by stronger ones. The point
  $x=x^*_{\{+--(+-)_\infty\}}$ is in the overlap at $\beta=1$, $J=1$,
  $h=0.74$ $\smash{(h_c^{(2)} < h < h_c^{(2a)}}$). In this figure $x$, all
  its predecessors up to $6$-fold application of $f_\pm^{-1}$ and all its
  successors up to $2$-fold application of $f_\pm$ are shown. The points are
  the predecessors/successors whereas the connecting lines illustrate which
  point is mapped onto which.  Note that the backward trajectory of $x$
  branches each time a point is in the overlap whereas the predecessor is
  unique if the overlap is not touched.  The dashed line connects points of
  the orbit $\{+--(+-)_\infty\}$ which is an offshoot of the $\{+-\}$ orbit
  and therefore has the singularity $\alpha_{\{+-\}}\approx 0.982$.  The
  solid lines connect points of two other orbits which meet at $y \in O$,
  both carrying the generic singularity $\overline{\alpha} = \overline{D}_p-1
  \approx -0.057$. At $x$ the weaker singularity $\alpha_{\{+-\}}$ is
  superseded by $\overline{\alpha}$. The offshoots emerging from $x$ all have
  the stronger singularity $\overline{\alpha}$.\label{fig4}}
\end{figure}

In the previous subsections the condition that no point of the orbits under
con\-si\-de\-ra\-tion is in the overlap $O$ was essential for the calculation
of their singularity. In this section we investigate how the singularity of
orbits is affected by the overlap $O$.  Tuning the parameter $h$ changes $O$
as well as the location of the orbits. If (at least) one point $x$ of a
periodic orbit is in $O$ this point has two predecessors and there are two
terms in the Frobenius-Perron equation (\ref{frobenius}) at this $x$. The two
terms contribute singularities $\alpha_1$ and $\alpha_2$, i.e.
$\mu_1:=\mu_\infty \big(B_\varepsilon(f_-^{-1}(x))\big) \sim
\varepsilon^{\alpha_1}$ and $\mu_2:=\mu_\infty
\big(B_\varepsilon(f_+^{-1}(x))\big) \sim \varepsilon^{\alpha_2}$.
Therefore, the singularity at $x$ will be $\inf \{\alpha_1, \alpha_2\}$,
since $\mu_\infty\big(B_\varepsilon(x)\big) \sim \mu_1+\mu_2 \sim
\varepsilon^{\inf\{\alpha_1, \alpha_2\}}$.  The mechanism is illustrated in
figure \ref{fig4} using the example of an orbit which will be important in
the next subsection.  In the case that the original singularity is stronger
than or equal to the additionally contributed one, there are no
consequences.  In the case that the singularity is rather weak though, a weak
singularity is replaced by a stronger singularity. In fact, the new
singularity at the maximal value of $h$ for which $x$ is in $O$ is always
rather strong as it stems from $x^*_\pm$ where $A'$ is small. The change in
the singularity may have a major impact on the $D_q$-spectrum
especially if some or all of the weak but somewhat stronger singularities
have already vanished. 

\begin{figure}
\begin{center}
\psfrag{l1}{\small $x^*_{-}$}
\psfrag{l2}{\small $x^*_{\{-+\}}$}
\psfrag{l3}{\small $f_+(x^*_-)$}
\psfrag{l4}{\small $f_-(x^*_+)$}
\psfrag{l5}{\small $x^*_{\{+-\}}$}
\psfrag{l6}{\small $x^*_{+}$}
\psfrag{l7}{\small $f_-$}
\psfrag{l8}{\small $f_+$}
\psfrag{l9}{\small $f_+$}
\psfrag{l10}{$f_-$}
\epsfig{file=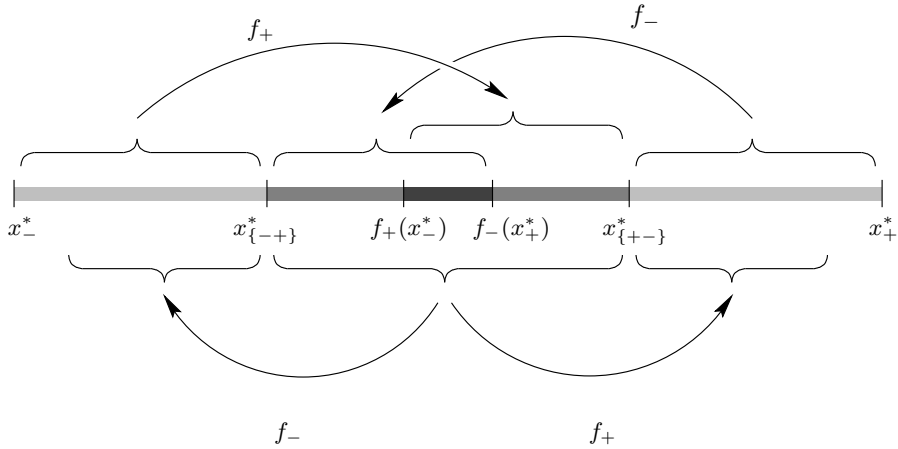, width=0.9\textwidth}
\end{center}
\caption{Mapping of subintervals of $I=[x^*_-,x^*_+]$ under $f_+$ and $f_-$
  elucidating the importance of the interval $[x^*_{\{-+\}},x^*_{\{+-\}}]$.
  The latter is mapped onto parts of the outer intervals
  $[x^*_-,x^*_{\{-+\}}]$ and $[x^*_{\{+-\}},x^*_+]$ under $f_-$ and $f_+$
  while these outer intervals are itself preimages of parts of
  $[x^*_{\{-+\}},x^*_{\{+-\}}]$. The points in the overlap
  $[f_+(x^*_-),f_-(x^*_+)]$ have two predecessors, one stemming from the left
  under $f_+$ and one from the right under $f_-$ (all in the case of moderate
  overlap).}
\label{fig5}
\end{figure}

A special role in the mechanism described above is played by the
$1$-orbits $\{+\}$ and $\{-\}$, because they never touch $O$,
and the $2$-orbit $\{+-\}$. At moderate overlap (small $|O|$) we have the
situation illustrated in figure \ref{fig5}.  Since $x^*_{\{+-\}}$ is mapped
to $x^*_{\{-+\}}$ and vice versa and because $f_\sigma$ is monotone, all
points to the right of $x^*_{\{+-\}}$ are mapped to the right of
$x^*_{\{-+\}}$ and all points to left of $x^*_{\{-+\}}$ are mapped to points
left of $x^*_{\{+-\}}$. Therefore, any periodic $n$-orbit with $n>2$ must have
at least one point inside $[x^*_{\{-+\}}, x^*_{\{+-\}}]$. Hence, the $\{+-\}$
orbit is the last periodic orbit to be reached by $O$.


The $\{+-\}$ orbit and its offshoots carry a very weak singularity as the
$\{+-\}$ orbit always stays in regions with comparably large $A'$, cf.
equation (\ref{alpha}). For $\beta$ and $J$ in the vicinity of $\beta=J=1$
one can show that the orbit $\{+-\}$ has even the weakest singularity of all
periodic orbits \cite{heiko}.  Because of its weak singularity and the fact
that all other periodic orbits and their offshoots are reached by $O$ before
$x^*_{\{+-\}}$ is reached, the $\{+-\}$ orbit and its offshoots practically
solely determine all $D_q$ with $q < 0$ if $h$ is such that $O$ has nearly
reached $x^*_{\{+-\}}$.  If $O$ includes $x^*_{\{+-\}}$, i.e.  $f_-(x^*_+)
\geq x^*_{\{+-\}}$, $x^*_{\{+-\}}$ has additionally
to the preimage $f_+^{-1}(x^*_{\{+-\}})$ the preimage
$f_-^{-1}(x^*_{\{+-\}})$. (The same applies to $x^*_{\{-+\}}$, of course.) The
additional preimages contribute a stronger singularity than the original weak
singularity of the $\{+-\}$ orbit. Thus, the weak singularity of the $\{+-\}$
orbit and its offshoots is superseded and, as all other periodic orbits have been reached by
$O$ before, all $D_q$ with negative $q$ have collapsed to $D_q =1$ at this
point.  The critical value $h_c^{(2)}$ at which the collapse takes place is
therefore given by the condition
\begin{align}
f_-(x^*_+) = x^*_{\{+-\}}.
\end{align}

So far we have only discussed periodic orbits and their offshoots.  Other
non-periodic orbits do not play a major role because they generically have
the rather strong singularity $\overline{\alpha}= \overline{D}_p -1 = D_1-1$ and also
generically have points inside $[x^*_{\{-+\}}, x^*_{\{+-\}}]$ such that they
are reached by $O$ before the $\{+-\}$ orbit is reached.

\begin{figure}[t]
  \psfrag{0}{\small $0$}
  \psfrag{0.2}{\small $0.2$}
  \psfrag{0.4}{\small $0.4$}
  \psfrag{0.5}{\small $0.5$}
  \psfrag{0.6}{\small $0.6$}
  \psfrag{0.8}{\small $0.8$}
  \psfrag{1}{\small $1$}
  \psfrag{1.2}{\small $1.2$}
    \psfrag{1.4}{\small $1.4$}
  \psfrag{1.6}{\small $1.6$}
  \psfrag{1.8}{\small $1.8$}
  \psfrag{1.5}{\small $1.5$}
  \psfrag{2}{\small $2$}
\begin{center}
\setlength{\unitlength}{\textwidth}
\begin{picture}(1,0.46)
\put(0.05,0.43){\bf a)}
\put(0.6,0.43){\bf b)}
  \put(-0.05,0){\epsfig{file=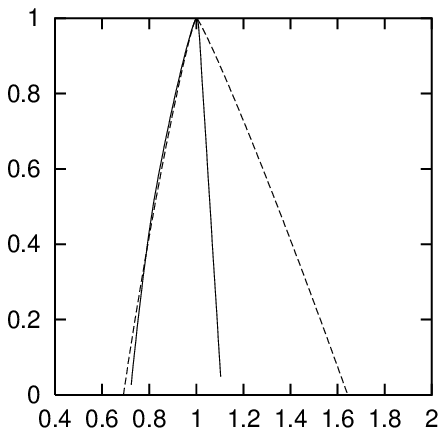, width=0.62\textwidth}}
  \put(0.5,0){\epsfig{file=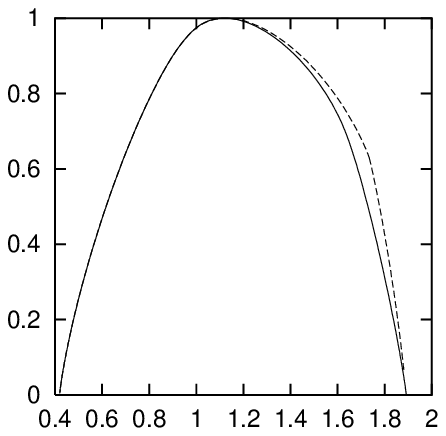, width= 0.62\textwidth}}
\end{picture}
\end{center}  
\caption{
  Collapse of the right part of the $f(\alpha)$-spectrum at $h_c^{(2)}$ and
  $h_c^{(2a)}$. In (a) the spectra at $h=
  0.49385 > h_c^{(2)}$ (dashed line) and $h= 0.4938 < h_c^{(2)}$
  (solid line) are shown. In (b) the spectra are at $h=0.8136 >
  h_c^{(2a)}$ (dashed line) and $h=0.8128 < h_c^{(2a)}$ (solid line).
  The spectra were obtained by a numerical Legendre transform of the
  corresponding $D_q$-spectra.  These were generated with the same
  algorithm as in figure \ref{fig2} with a recursion depth of 21 ($\beta=1$,
  $J=1$).
\label{fig6}}
\end{figure}   


 After the collapse the right part of the $f(\alpha)$-spectrum
of the invariant measure has vanished since the weaker negative singularities
have all been superseded by stronger ones, cf. figure \ref{fig6}a. A similar
collapse of parts of the multifractal spectrum has previously been observed
in the superposition of equal-scale \cite{radons1} and multi-scale
\cite{radons2,stoop1} Cantor sets showing that effects of this type appear in
a wide variety of applications.

Let us now determine $h_c^{(2)}$ explicitly. In a first step we need explicit
expressions for $x^*_+= - x^*_-$ and $x^*_{\{+-\}} = - x^*_{\{-+\}}$.
The fixed point $x^*_+$ is defined by $f_+(x^*_+) = x^*_+$. With the notation
$z=e^{2\beta x^*_+}$ this yields the equation
\begin{align}
& z^2 - (e^{2\beta J}(e^{2\beta h}-1))z - e^{2\beta h} = 0
\end{align}
with the solution 
\begin{align}
 & x^*_+= \frac{1}{2\beta} \ln \big( K + \sqrt{K^2 + e^{2\beta
h}}\big)
\end{align}
where $K=e^{2\beta J}(e^{2\beta h} -1)/2$.
To obtain $x^*_{\{+-\}}$ we exploit $x^*_{\{+-\}} = -x^*_{\{-+\}}= -
f_-(x^*_{\{+-\}})$. With $z= e^{2\beta x^*_{\{+-\}}}$ this yields
\begin{align}
z^2 + e^{-2\beta J}(e^{2\beta h}-1) z - e^{2\beta h} = 0 
\end{align}
and therefore
\begin{align}
 x^*_{\{+-\}} = \frac{1}{2\beta} \ln\big( \tilde{K} +
\sqrt{\tilde{K}^2 +e^{2\beta h}} \big)
\end{align}
with $\tilde{K}= e^{-2\beta J}(e^{2\beta h}-1)/2$.\begin{figure}
\begin{center}
{\psfrag{0}{0}
\psfrag{0.5}{0.5}
\psfrag{1}{1}
\psfrag{1.5}{1.5}
\psfrag{2}{2}
\psfrag{3}{3}
\psfrag{label1}{\hspace{1cm} $\beta$}
\psfrag{label2}{$J$}
\psfrag{label3}{$h_c^{(2)}$}
\epsfig{file=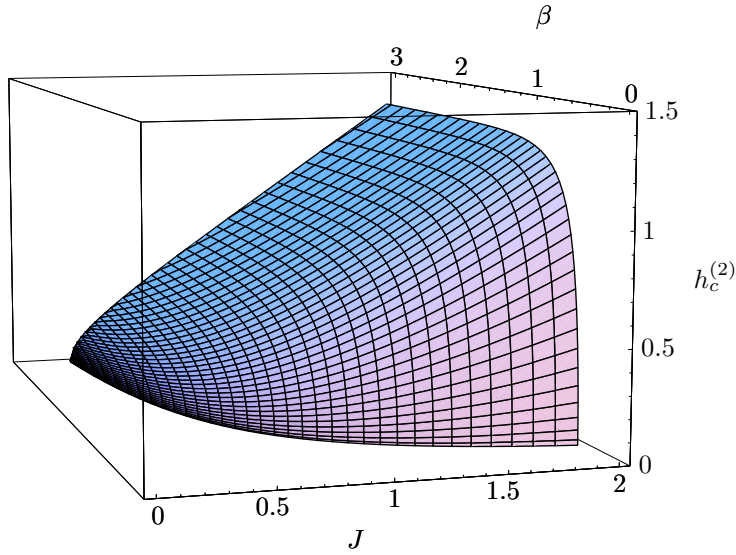, width=0.8\textwidth}
}
\end{center}
\caption{Surface plot of the critical value $h_c^{(2)}$ against
  $\beta$ and $J$. The plot was obtained from (\ref{hc2form}).
  }
\label{fig7}
\end{figure}

 The equation
$f_-(x^*_+) = x^*_{\{+-\}}$ is equivalent to $x^*_+= x^*_{\{+-\}} + 2h$. With
the explicit expressions for $x^*_+$ and $x^*_{\{+-\}}$ we get
\begin{align}
2 \big(\cosh(2\beta h_c^{(2)})\big)^3 + 3 \big(\cosh(2\beta h_c^{(2)})\big)^2
= \big(\cosh(2 \beta J)\big)^2 .
\end{align}
This equation has exactly one real solution for $\cosh \big(2 \beta h_c^{(2)}\big)$
resulting in
\begin{align}
  h_c^{(2)}= \frac{1}{2\beta} \arcosh\big(\cosh(\textstyle \frac{4}{3}\beta J)-
  \textstyle \frac{1}{2}\big) . \label{hc2form}
\end{align}
The phase diagram for the transition at
$h_c^{(2)}$ is shown in figure \ref{fig7}. One clearly
sees that there is a critical line $\beta(J)$ such that there is no
phase transition possible for all $\beta < \beta(J)$. This line is
given by the condition $h_c^{(2)}= 0$ corresponding to
\begin{align}
\beta(J)=\frac{3}{4J} \arcosh({\textstyle \frac{3}{2}}) = \frac{1}{2 J} \ln \big(2 + \sqrt{5} \big) . \label{hc2bound}
\end{align}
With (\ref{hc2form}) and (\ref{hc2bound}) we thus have a complete analytical
understanding of the occurrence and the position of the transition at
$h_c^{(2)}$.

\subsection{Transition of the measure at $h_c^{(2a)}$}\label{hc2asec}
We have seen in the last subsection that the sharp drop of all $D_q$ with
negative $q$ at $h_c^{(2)}$ can be explained through the analysis of the
properties of the $\{+-\}$ orbit which is on the one hand an orbit with very
weak singularity and on the other hand the last periodic orbit (with
decreasing $h$) to be affected by th overlap $O$. To understand the smaller
drop at $h_c^{(2a)}$ it is therefore natural to look for periodic orbits with
weak singularity which also are affected by the overlap relatively late. We
will now argue that the family of periodic orbits of the form
$\{{+\!-\!-}(+-)_n\}$, $n \in \N$ fulfills these conditions and that their
change in singularity indeed causes the transition at $h_c^{(2a)}$. To
present the precise argument we need to investigate some properties of this
family of orbits.

For large $n$ most points of the $\{{+\!-\!-}(+-)_n\}$ orbits are close to
the points of the $\{+-\}$ orbit. Thus, having (\ref{alpha}) in mind, the
singularities of these orbits have a similar strength as singularity of the
$\{+-\}$ orbit, i.e. they are weak, provided that no point of the orbits is
in the overlap $O$. The larger $n$ the larger is the fraction of points of
the corresponding orbit which are close to $x^*_{\{+-\}}$ and $x^*_{\{-+\}}$.
Therefore, the singularities get weaker with growing $n$.

Concerning the position of the orbits $\{{+\!-\!-}(+-)_n\}$ we first note that the
points $z_n := x^*_{\{{+--}(+-)_n\}}$ are for each of the orbits
$\{{+\!-\!-}(+-)_n\}$ the closest points to the overlap $O$. As $f_{\{{+--}(+-)_n\}}
(x_0^{\sst (n)}) \to x^*_{\{{+--}(+-)_\infty\}}$ for any choice of initial
points $x_0^{\sst (n)} \in I$ and $n \to \infty$, we get with $x_0^{\sst (n)}:= z_n=
x^*_{\{{+--}(+-)_n\}}$ that $z_n \to x^*_{\{{+--}(+-)_\infty\}}$ for $n \to \infty$.
Furthermore, it is an easy exercise to check that $z_n$ is monotonously
growing with $n$.

Taking these properties together we conclude that each of the periodic orbits
of the form $\{{+\!-\!-}(+-)_n\}$ is affected by the overlap as soon as $z_n$
is in $O$ and that the orbits with weaker singularity (those with large $n$)
are affected by the overlap later than the orbits with somewhat stronger
singularity. Finally, as the $z_n$ converge to $x^*_{\{{+--}(+-)_\infty\}}$,
there are still countably infinitely many orbits $\{{+\!-\!-}(+-)_n\}$, $n
\geq N$ with some $N \in \N$ left which are not affected by the overlap as
long as $x^*_{\{{+--}(+-)_\infty\}}$ is not in $O$. As soon as
$x^*_{\{{+--}(+-)_\infty\}}$ gets into the overlap $O$, all orbits
$\{{+\!-\!-}(+-)_n\}$ are in the overlap and their singularity therefore
superseded.  The critical field strength $h_c^{(2a)}$ is thus determined by
the condition \begin{figure}
\begin{center}
{\psfrag{0.95}{$h_c^{(2a)}$}
\psfrag{0.9}{0.9}
\psfrag{0.85}{0.85}
\psfrag{0.8}{0.8}
\psfrag{0.75}{0.75}
\psfrag{0.7}{0.7}
\psfrag{0.65}{0.65}
\psfrag{0.6}{0.6}
\psfrag{0.55}{0.55}
\psfrag{1}{1}
\psfrag{1.1}{1.1}
\psfrag{1.2}{1.2}
\psfrag{1.3}{1.3}
\psfrag{1.4}{$\beta$}
\epsfig{file=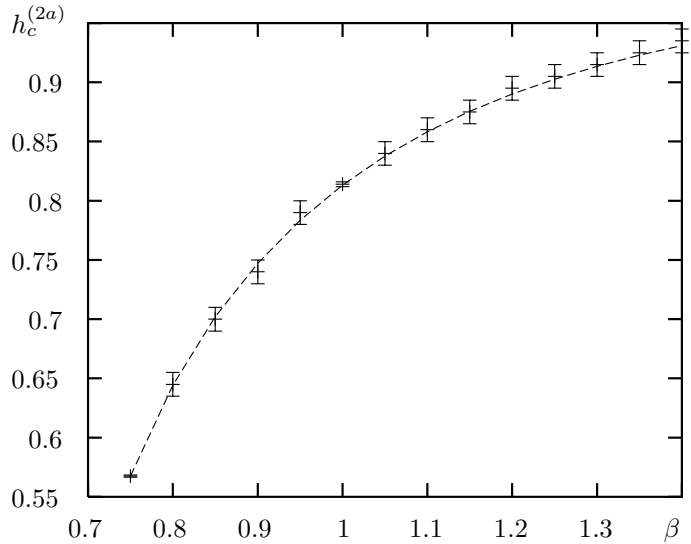, width=0.8\textwidth}
}
\end{center}
\caption{$h_c^{(2a)}$ obtained from the analytical condition (\ref{hc2acond})
  (dashed line)
  compared with the location of the small drops in $D_q$ 
   (error-bars), cf. figure \ref{fig2}. The errors are estimated and mainly 
  due to the errors in the determination of the location of the drops
  in the numerically obtained $D_q$-spectra ($J=1$).}
\label{fig8}
\end{figure}
%
\begin{align}
   f_-(x^*_+) = x^*_{+--(+-)_\infty}= f_{\{+-\}}(x^*_{\{-+\}}) .
  \label{hc2acond}
\end{align}
This criterion is in perfect agreement with the numerically obtained
positions of the small drop in the $D_q$-spectrum for $q=-1, -2, -3$, cf.
figure \ref{fig8}. The transition is also visible in the
$f(\alpha)$-spectrum. For $h \to h_c^{(2a)}+0$ a cusp develops and the
spectrum collapses to a smooth form again at $h_c^{(2a)}$, cf. figure
\ref{fig6}b.

So far we only addressed the periodic orbits $\{{+\!-\!-}(+-)_n\}$ neglecting their
offshoots. Each such orbit has countably many offshoots. From these offshoots
those originating from points left of $O$ and containing exclusively
additional $-$ as well as those originating from points to the right of $O$
and only containing additional $+$ are also not affected by the overlap as
long as the corresponding periodic orbit is not. Thus, at $h_c^{(2a)}$
countably infinitely many periodic orbits with each countably many offshoots
of the described form vanish at once. This explains why this
transition is visible in the $D_q$-spectrum in contrast to
events of single orbits (and their offshoots) being affected by the overlap
at any value of $h$.

The fact that the transition is not visible in $D_q$ with large negative $q$
is also easily understood in this framework. As the singularities of all
$\{{+\!-\!-}(+-)_n\}$ orbits are similar to but somewhat stronger than the
singularity of the $\{+-\}$ orbit, the $\{+-\}$ orbit and its offshoots
dominate $D_q$ for large negative $q$ and the transition is not visible.

Please note the symmetry of the system
which allows the same reasoning with the `opposite' orbits $\{-\!+\!+(-+)_n\}$
resulting in an equivalent result.

\begin{figure}
\psfrag{-10}{$\nts{17}$ \raisebox{-0.08cm}{\parbox{1cm}{\hfill \tiny$-10$}}}
\psfrag{-8}{$\nts{18}$ \raisebox{-0.08cm}{\parbox{1cm}{\hfill \tiny$-8$}}}
\psfrag{-7}{$\nts{18}$ \raisebox{-0.08cm}{\parbox{1cm}{\hfill \tiny$-7$}}}
\psfrag{-6}{$\nts{18}$ \raisebox{-0.08cm}{\parbox{1cm}{\hfill \tiny$-6$}}}
\psfrag{-5}{$\nts{18}$ \raisebox{-0.08cm}{\parbox{1cm}{\hfill \tiny$-5$}}}
\psfrag{-4}{$\nts{18}$ \raisebox{-0.08cm}{\parbox{1cm}{\hfill \tiny$-4$}}}
\psfrag{-3}{$\nts{18}$ \raisebox{-0.08cm}{\parbox{1cm}{\hfill \tiny$-3$}}}
\psfrag{-2}{$\nts{18}$ \raisebox{-0.08cm}{\parbox{1cm}{\hfill \tiny$-2$}}}
\psfrag{-1}{$\nts{18}$ \raisebox{-0.08cm}{\parbox{1cm}{\hfill \tiny$-1$}}}
\psfrag{0}{$\nts{18}$ \raisebox{-0.08cm}{\parbox{1cm}{\hfill \tiny$0$}}}
\psfrag{1}{$\nts{18}$ \raisebox{-0.08cm}{\parbox{1cm}{\hfill \tiny$1$}}}
\psfrag{2}{$\nts{18}$ \raisebox{-0.08cm}{\parbox{1cm}{\hfill \tiny$2$}}}
\psfrag{3}{$\nts{18}$ \raisebox{-0.08cm}{\parbox{1cm}{\hfill \tiny$3$}}}
\psfrag{4}{$\nts{18}$ \raisebox{-0.08cm}{\parbox{1cm}{\hfill \tiny$4$}}}
\psfrag{5}{$\nts{18}$ \raisebox{-0.08cm}{\parbox{1cm}{\hfill \tiny$5$}}}
\psfrag{6}{$\nts{18}$ \raisebox{-0.08cm}{\parbox{1cm}{\hfill \tiny$6$}}}
\psfrag{7}{$\nts{18}$ \raisebox{-0.08cm}{\parbox{1cm}{\hfill \tiny$7$}}}
\psfrag{xlabel}{\raisebox{-0.18cm}{\small $\nts{14} -\ln (x^*_{+}-x)$}}
\psfrag{ylabel}{\raisebox{0.18cm}{\small $\nts{10} \ln (p_n)(x)$}}
\parbox{0.48\textwidth}{{\bf a)} $h= 0.85 > h_c^{(2a)}$} \hfill
\parbox{0.48\textwidth}{{\bf b)} $h= 0.7 < h_c^{(2a)}$} \\
\epsfig{file=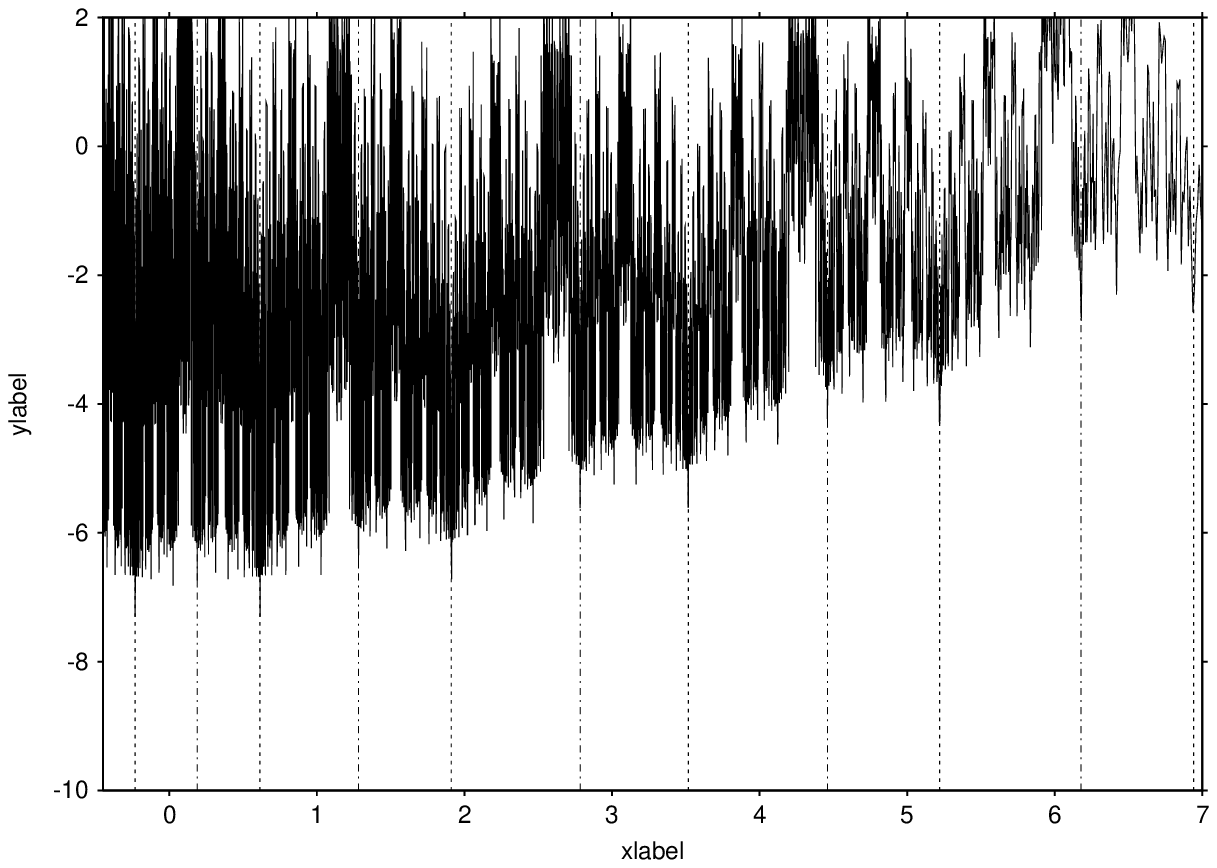, width=0.48\textwidth} \hfill
\epsfig{file=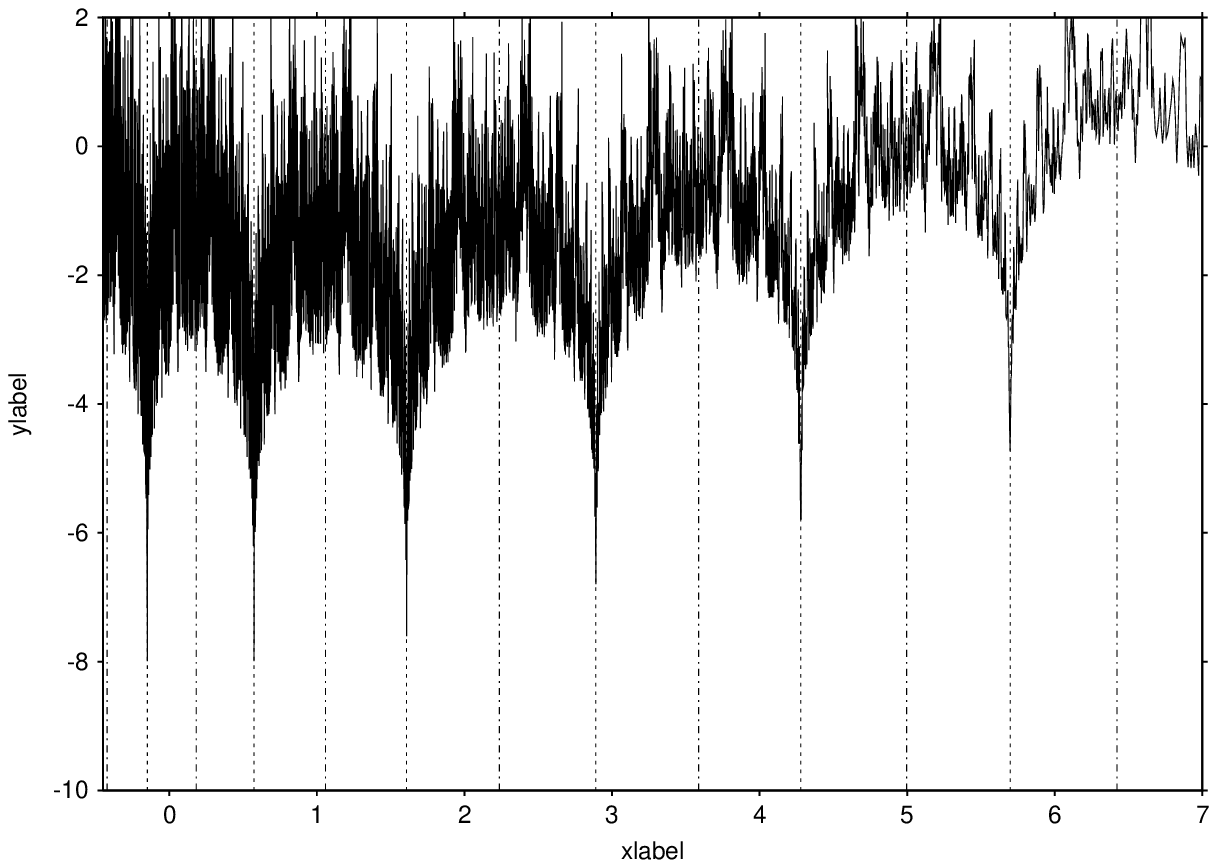, width=0.48\textwidth} \\[0.25cm]
\parbox{0.48\textwidth}{{\bf c)} $h\approx 0.5436 > h_c^{(2)}$} \hfill
\parbox{0.48\textwidth}{{\bf d)} $h= 0.48 < h_c^{(2)}$} 
\epsfig{file=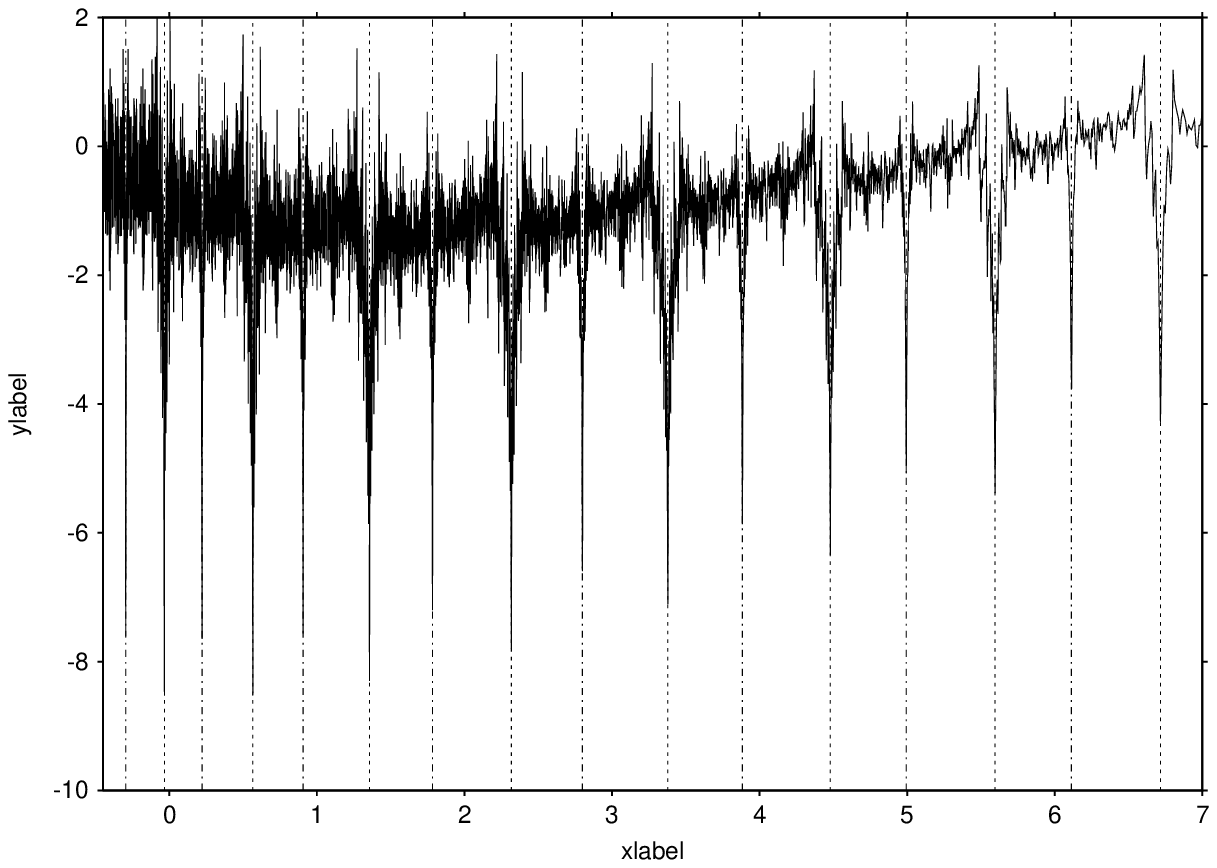, width=0.48\textwidth} \hfill
\epsfig{file=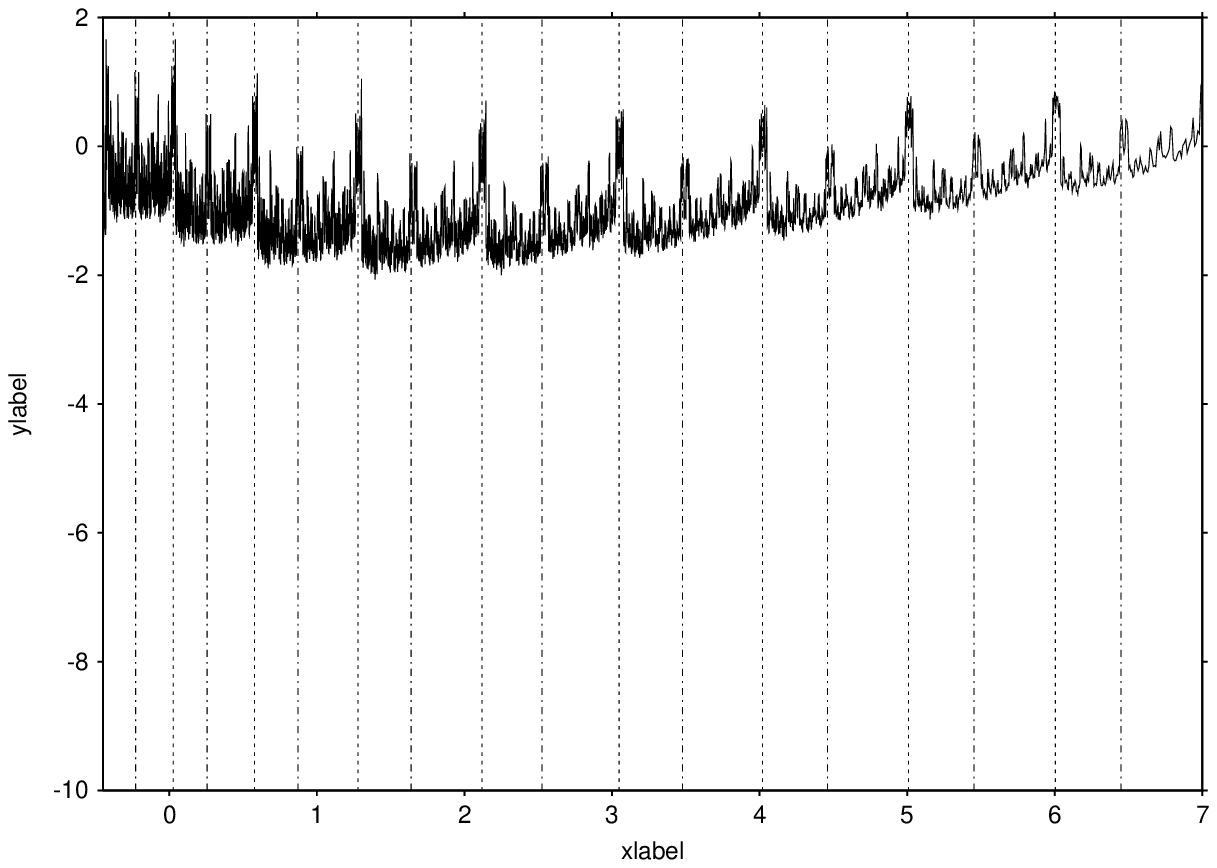, width=0.48\textwidth}
\caption{Double logarithmic plot of $p_n$  
  generated from the band structure after $n=21$ iterations. The negative peaks
  correspond to deep cuts in $p_n$. The dotted lines mark the position of
  $x^*_{\{(+)_k(+-)_\infty\}}$, i.e. the points of the $\{+-\}$ orbit and
  its offshoots, and the dash-dotted lines the positions of
  $x^*_{\{(+)_k+--(+-)_\infty\}}$, i.e. the points of the
  $\{+\!-\!-(+-)_\infty\}$ orbit and its offshoots. In (a) for $h > h_c^{(2a)}$
  deep cuts are visible at all marked positions whereas in (b) for $h <
  h_c^{(2a)}$ the deep cuts at $x^*_{\{(+)_k+--(+-)_\infty\}}$ have
  vanished. They reappear in (c) at a critical field strength $h \approx
  0.5436$ as is discussed in the text whereas in (d) for $h
  < h_c^{(2)}$ all deep cuts have vanished
  ($\beta=1$, $J=1$).} \label{fig9} 
\end{figure}


Even though the effect is due to the periodic orbits $\{{+\!-\!-}(+-)_n\}$ and
their offshoots, the orbit entering into condition (\ref{hc2acond}) is an
offshoot of the $\{+-\}$ orbit, namely $\{{+\!-\!-}(+-)_\infty\}$. This
orbit thus seems to play a similar role as the $\{+-\}$ orbit for the
transition at $h_c^{(2)}$. For illustration we have generated an
approximation of $p_n$ and calculated the position of the points of
the $\{+-\}$ orbit and its offshoots of the form $\{(+)_n(+-)_\infty\}$ (dotted
lines in figure \ref{fig9}) as well as the position of the points of the
$\{{+\!-\!-}(+-)_\infty\}$ orbit and its offshoots of the form
$\{(+)_n+\!-\!-(+-)_\infty\}$ (dashed lines in figure
\ref{fig9}). It is obvious that there is a bunch of weak singularities in
the vicinity of $x^*_{\{{+--}(+-)_n\}}$ and in the vicinity of the points of the
offshoots which all vanish when decreasing $h$ below $\smash{h_c^{(2a)}}$, cf. figure
\ref{fig9}a and b.

The attentive reader will have noticed that we have argued that the weak
singularity of an orbit which is touched by the overlap $O$ for some $h_0$
but not for $h > h_0$ will be superseded by the strong singularity of the
$\{+\}$ or the $\{-\}$ orbit. For smaller $h$ in the generic
case the additionally contributed singularity will be $D_1-1$ and thus also
rather large. It is not excluded though that two weak singularities are combined
resulting in a weak singularity even though the orbit is in the overlap. A
prominent example of this effect is the value of $h$ for which $x^*_{\{{+--}(+-)_\infty\}} =
x^*_{\{-++(-+)_\infty\}}= 0$. At this value of $h$ new deep cuts in the
approximated measure density appear at $x^*_{\{{+--}(+-)_\infty\}}$ and its
offshoots, cf.~figure \ref{fig9}c. The effect on the $D_q$-spectrum is
negligible though as this is a rare event only affecting a single orbit
and its offshoots at a given $h$.

To summarize, the crucial feature for a visible transition in the
$D_q$-spectrum is that a non-negligible fraction of orbits with weak
singularities is affected by the overlap at one sharp critical value $h_c$
resulting in a drop of $D_q$ with negative $q$. It is not excluded that there
are more transitions of this type which might become observable with further
increase in numerical accuracy in the future. Our arguments hint to the
conjecture that these transitions, if existent, should take place at $h >
h_c^{(2a)}$.

\begin{figure}
\setlength{\unitlength}{\textwidth}
\parbox{\textwidth}{\rule{0cm}{0.5\textwidth}
\begin{center}
\begin{picture}(0.8,0)
\put(0,0){
{\psfrag{0}{0}
\psfrag{0.2}{0.2}
\psfrag{0.4}{0.4}
\psfrag{0.5}{0.5}
\psfrag{0.6}{0.6}
\psfrag{0.8}{0.8}
\psfrag{1}{1}
\psfrag{1.2}{}
\psfrag{1.5}{1.5}
\psfrag{2}{}
\epsfig{file=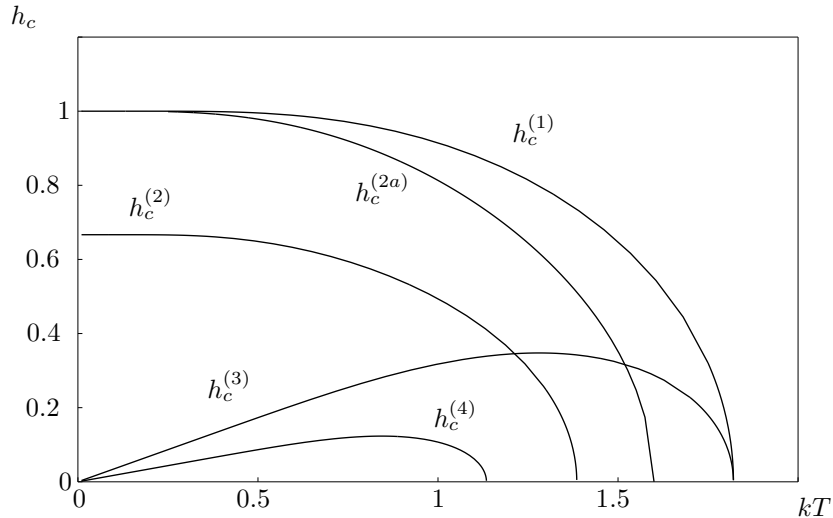, width=0.8\textwidth}
}}
\put(0,0.5){$h_c$}
\put(0.8,0){$kT$}
\put(0.51,0.38){$h_c^{(1)}$}
\put(0.35,0.32){$h_c^{(2a)}$}
\put(0.12,0.305){$h_c^{(2)}$}
\put(0.2,0.12){$h_c^{(3)}$}
\put(0.43,0.09){$h_c^{(4)}$}
\end{picture}
\end{center}
}
\caption{Phase diagram for the transitions of the invariant
  measure derived from the random field Ising model. The $h_c^{(n)}$ are the
  critical random field strengths defined in the text.  The remarkable fact
  that the lines of $h_c^{(2)}$ and $h_c^{(3)}$ as well as the lines of
  $h_c^{(2a)}$ and $h_c^{(3)}$ intersect shows that there is a variety of
  different scenarios depending on the choice of $kT$. For example at
  $kt=1.3$, the transition at $h_c^{(2)}$ precedes the one at $h_c^{(3)}$
  while the transition at $h_c^{(4)}$ is non-existent ($J=1$).}
\label{fig10}
\end{figure}


The formulae given in the last two subsections allow to draw a
phase diagram for the phase transitions in the $D_q$-spectrum of the invariant
measure which have been observed so far, cf. figure
\ref{fig10}.  It should be emphasized that these `phase transitions' are not
phase transitions of physical quantities like the (local) magnetisation or
the Edwards-Anderson parameter.

\section{Bounds on the {\boldmath$D_q$}-spectrum} \label{boundsec}
Let in the following $\mu$ denote the invariant measure. As mentioned above there are natural bounds on $D_q$. They are
induced by
\begin{align}
  \mu_{(x)}^q \leq \sum_{i} \mu_i^q \label{basicineq}
\end{align}
where $x$ is some point in the support of $\mu$ and $\mu_{(x)}$ is the
measure of the box containing $x$. Let us now first consider the case $q <
0$. In this case 
inequality (\ref{basicineq}) immediately yields
\begin{align}
  D_q \geq \frac{q}{q-1} \lim_{\varepsilon \to 0} \frac{\ln( \mu_{(x)})}{\ln
  \varepsilon} .
\end{align}
The limit on the right hand side is the pointwise dimension at $x$. If we
choose $x= x^*_+$ we can calculate the pointwise dimension at $x$ for any $h,
T \not= 1, J$
using (\ref{alpha}). This gives the lower bound
\begin{align}
  D_q \geq \frac{q}{1-q} \frac{\ln 2}{\ln (A'(x^*_+))} . \label{lowboundm1}
\end{align}
In the region $h > h_c^{(2)}$ we also can calculate the pointwise dimension
at $x^*_{\{+-\}}$. With $x= x^*_{\{+-\}}$ we get the bound 
\begin{align}
   D_q \geq \frac{q}{1-q} \frac{\ln 2}{\ln (A'(x^*_{\{+-\}}))} . \label{lowboundm2}
\end{align} 
These lower bounds are shown as solid lines in figure \ref{fig2}.

Whenever we know that the invariant measure scales most weakly at some point $x$
of its support and we can calculate the pointwise dimension at $x$, we
also can give an upper bound induced by
\begin{align}
  \sum_i \mu_i^q \leq N \mu_{(x)}^q . \label{abcd}
\end{align}
In this $N$ is the number of boxes and is essentially proportional to
$1/\varepsilon$. In the region $h < h_c^{(3)}$ the measure
is assumed to scale most weakly at $x^*_+$ (and $x^*_-$) such that inserting
(\ref{abcd}) into (\ref{efgh}) we obtain
the upper bound 
\begin{align}
  D_q \leq \frac{1}{1-q} \Big(1+\frac{q \ln 2}{\ln(A'(x^*_+))}\Big).
\end{align}
In the region $h > h_c^{(2)}$ the scaling is assumed to be weakest at
$x^*_{\{+-\}}$ (and $x^*_{\{-+\}}$) yielding 
\begin{align}
  D_q \leq \frac{1}{1-q} \Big(1+\frac{q \ln 2}{\ln(A'(x^*_{\{+-\}}))}\Big).
\end{align}
Obviously the upper and lower bounds converge against each other in both
regions as $q \to -\infty$ such that we get the explicit expressions
\begin{align}
  D_{-\infty}= \left\{ \begin{array}{ll}
      -\frac{\ln 2}{\ln(A'(x^*_+))} & h < h_c^{(3)} \\
      -\frac{\ln 2}{\ln(A'(x^*_{\{+-\}}))} & h > h_c^{(2)} \label{Dminfeq}
    \end{array} \right. .
\end{align}
The expression for the region $h < h_c^{(3)}$ was already given in \cite{behn7}
whereas the expression for the region $h > h_c^{(2)}$ needs our analysis of
the singularity of orbits in section \ref{singorbitsec}. These upper bounds are
in general not as good as the lower ones found above and are not shown in
figure \ref{fig2}.

In the case $q > 0$ all considerations remain valid for the bounds based on
(\ref{basicineq}) which yields by use of (\ref{alpha}) the upper bounds
\begin{align}
  D_q \leq \frac{q}{1-q} \frac{\ln 2}{\ln(A'(x^*_+))} \label{upboundp1}
\end{align}
for any $h$ and the sharper condition
\begin{align}
  D_q \leq \frac{q}{1-q} \frac{\ln 2}{\ln(A'(x^*_{\{+-\}}))} \label{upboundp2}
\end{align}
for $h > h_c^{(2)}$. These upper bounds (\ref{upboundp2}) are shown as dashed
lines in figure \ref{fig2}.

Lower bounds on $D_q$ can be obtained by considering
points $x$ at which the scaling is maximally strong. This is assumed to be
true for $x^*_+$ and $x^*_-$ in the region $h > h_c^{(3)}$. Therefore we get
\begin{align}
  D_q \geq \frac{1}{1-q} \Big( 1+ \frac{q \ln 2}{\ln(A'(x^*_+))} \Big) .
\end{align}
These lower bounds are not as good as the upper bounds (\ref{upboundp2}) and
are not shown in figure \ref{fig2}.
In the region $h > h_c^{(2)}$ the lower and upper bounds for $D_q$ converge
against a common value for $q \to \infty$ yielding 
\begin{align}
  D_\infty = -\frac{\ln2}{\ln(A'(x^*_+))}. \label{Dinfeq}
\end{align}
The expressions (\ref{Dminfeq}) and (\ref{Dinfeq}) generalize results
priorly obtained \cite{evangelou} for $h > h_c^{(1)}$.

Note that for $q < 0$ ($q > 0$) the upper (lower) bounds rest on the
assumption of minimal (maximal) scaling at $x^*_+$ ($x^*_-$) and
$x^*_{\{+-\}}$ ($x^*_{\{-+\}}$) whereas the lower (upper) bounds do not need
any additional assumptions. In figure \ref{fig2} only the lower (upper)
bounds are shown.

The fact that for $q \leq -2$ ($q \geq 4$) the lower (upper) bounds 
are more or less identical with the numerical data (cf. figure \ref{fig2})
shows that in these cases the generalized dimensions $D_q$ are solely
determined by the scaling at certain points, i.e. at $x^*_+$ and $x^*_-$ for
$h < h_c^{(3)}$ and at $x^*_{\{+-\}}$ for $h > h_c^{(2)}$. This reinforces our
previous argument why the small drop in the $D_q$-spectrum at $h_c^{(2a)}$ is
not visible in numerical data for $ q < -3$.

\section{Concluding remarks}\label{concludesec}

By generalizing arguments, hitherto mainly applied to fixed points, to orbits
we have been able to calculate the singularity of all periodic orbits not
touching the overlap $O$ and the generic ($\mu$-a.s.) singularity of orbits
in the non-overlapping case. We then investigated the effects of a non-void
overlap on the singularity of orbits. While being a relevant result in its
own right the knowledge of the singularity of orbits and their dependence on
touching the overlap or not also provided the explanation for two phase
transitions in the multifractal $D_q$-spectrum of the invariant measure.

RIFS with similar properties as the one discussed here also appear in a
variety of other contexts \cite{schmidt, halperin, barnes, behn7}.
In \cite{behn7} a similar phase transition in the $D_q$-spectrum already has
been observed. The explanation of the mechanism causing such transitions does
not crucially depend on the special choice of the RIFS we investigated
here.
Let us briefly discuss some obvious generalizations.

Dropping the restriction of symmetry of the random field distribution leads
to the loss of various symmetries of the orbit structure, e.g. $x^*_+ \neq
- x^*_-$ and $x^*_{\{+-\}} \neq - x^*_{\{-+\}}$, and of the resulting
calculational simplicity. The general orbit structure underlying our
explanations of transitions in the $D_q$-spectrum however
remains. Therefore, the transitions also exist in the non-symmetric case and
can be analysed in the same way as above in the symmetric case. Some
numerical results may be found \cite{behn6}. The explicit
calculations are far more complicated though.

\begin{figure}[t]
\begin{center}
  \psfrag{t1}{\footnotesize $x^*_1$}
  \psfrag{t3}{\footnotesize $x^*_2$}
  \psfrag{t4}{\footnotesize $x^*_{\{23\}}$}
  \psfrag{t5}{\footnotesize $O_{23}$}
  \psfrag{t6}{\footnotesize $x^*_{\{32\}}$}
  \psfrag{t7}{\footnotesize $x^*_4$}
  \psfrag{t8}{$I_1$}
  \psfrag{t9}{$I_2$}
  \psfrag{ta}{$I_3$}
  \psfrag{tb}{$I_4$}
  \psfrag{tc}{\footnotesize $x^*_3$}
  \epsfig{file=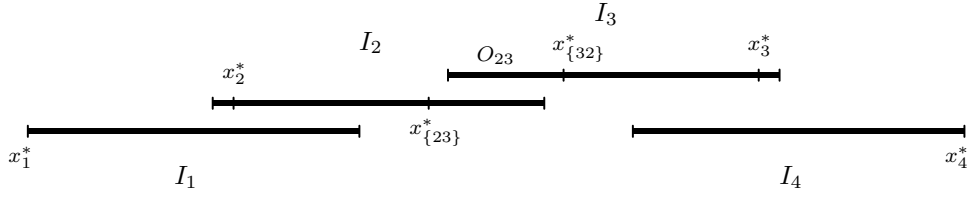, width=\textwidth}
\end{center}  
\caption{
  Sample situation of first order bands in the symmetric four Dirac mass case
  in the vicinity of a transition similar to the one at $h_c^{(2)}$ in the
  text.  If, as shown in the figure, $H$ is large enough, $x^*_{\{23\}}$ and
  $x^*_{\{32\}}$ are not included in $I_1$ and $I_2$ respectively. In this
  case $x^*_{\{23\}}$ has the unique predecessor $x^*_{\{32\}}$ and vice
  versa exactly like $x^*_{\{-+\}}$ and $x^*_{\{+-\}}$ in the dichotomous
  case. The situation is therefore identical to the situation in the
  dichotomous case and the transition takes place at $h= h_c^{(2)}$.
  ($\beta= 1$, $h=0.7$, $H= 1.8$)
\label{fig11}}
\end{figure}   

 Considering more complex random field distributions like
e.g.~$n$ (symmetrically or non-symmetrically distributed) Dirac masses
considerably complicates the system. The mechanisms causing phase transitions
in the $D_q$-spectra are nevertheless still the same. Let us briefly outline
the example of four symmetrically distributed Dirac masses located at
$\{-H,-h,h,H\}$ as the random field distribution, i.e. the RIFS $\{f_1, f_2,
f_3, f_4\}= \{A-H, A-h, A+h, A+H\}$ with $h < H \in \R^+$:
\begin{itemize}
  \item For any given $\beta$ a transition line in the $h$-$H$-plane
    exists between a region with $D_0=1$ and $D_0 < 1$ which is easily determined
    from the overlap conditions of the first order bands $I_j$, $j= 1, \ldots,
    4$. These conditions read $h \leq h_c^{(1)}$ and $f_3(x^*_4) \geq f_4(x^*_1)$,
    the second inequality being a condition on $H$ as well as $h$.
  \item The transitions depending on the scaling at the boundary of the
    support of $\mu_\infty$, i.e. on the scaling at $x^*_1$ and $x^*_4$,
    exist in the same way as those at $h_c^{(3)}$ and $h_c^{(4)}$ above. The
    transition conditions are $H = h_c^{(3)}$ and $H= h_c^{(4)}$
    respectively.
  \item Even though the orbit structure is more intricate than in the
    dichotomous case figure \ref{fig11} shows a situation in which a
    transition of the type of the one at $h_c^{(2)}$ above can take place. The
    transition occurs at $h= h_c^{(2)}$ provided $H$ is large enough for the
    bands $I_1$ and $I_2$ not to include $x^*_{\{23\}}$ or $x^*_{\{32\}}$,
    i.e. $f_1(x^*_4) < x^*_{\{23\}}$ or equivalently $f_4(x^*_1) >
    x^*_{\{32\}}$.
\end{itemize}
In summary, the types of transitions occurring in the $D_q$-spectrum are the
same as in the dichotomous case and it is possible to draw a diagram in the
$(\beta,h,H)$ parameter space similar to the simpler diagram in figure
\ref{fig10} which is the $h=H$ slice of this higher dimensional diagram.

In the same fashion the analytic tools developed in this paper in principal
allow to analyse the $D_q$-spectrum of the distribution of the
effective field of the RFIM for any discrete random field distribution.
Moreover, the exact form of the function $A$ is not crucial such that the
analysis can be performed as soon as the following features are present:
\begin{itemize}
\item A hyperbolic    RIFS $\{f_j\}$, $j=1, \ldots, n$ which -- in dependence on a
  control parameter $h$ -- has overlapping bands or not.
\item The functions $f_j$ are sufficiently smooth, monotonous and
  have com\-pa\-rab\-ly large derivatives at certain periodic orbits.
\item The conditions for the transitions that certain (periodic) orbits touch
  or do not touch the overlap can be fulfilled by tuning
  $h$.
\end{itemize}

The generalization to continuous random field distributions is not obvious as
the techniques used above can not directly be applied to this case. In
\cite{andelman} a numerical survey of the nature of the distribution function
$P_\infty$ for various random field distributions can be found. For discrete
random fields a transition between a devils's staircase corresponding to $D_0
< 1$ and a smooth function corresponding to $D_0 = 1$ is found as is
predicted analytically. For continuous random field distributions without
gaps $P_\infty$ always is smooth whereas gaps in the continuous random field
distribution define a scale above which (for certain parameters) $P_\infty$
resembles a devil's staircase and below which it is always smooth.

The generalization to more general lattices like the dichotomous symmetric
RFIM on the Bethe lattice (cf. \cite{bruinsma2, brz} and references therein)
is non-trivial. Transitions corresponding to those at $h_c^{(1)}$,
$h_c^{(3)}$ and $h_c^{(4)}$ above are present. The exact conditions
determining the critical field strengths are however not obvious and need
further careful investigation. The existence and properties of transitions of
the type of those at $h_c^{(2)}$ or $h_c^{(2a)}$ needs even more careful and
detailed analysis and must be deferred to further work. We stress however
that all these transitions in the $D_q$-spectrum take place far in the
contracting, paramagnetic regime of the physical phase diagram of the RFIM on
the Bethe lattice.

The bounds on the $D_q$-spectrum obtained in section \ref{boundsec} are also
of a quite general type and therefore applicable in a variety of contexts. As
figure \ref{fig2} shows, the bounds together with the explanation of all the
transitions discussed give a very good qualitative as well as quantitative
understanding of the $D_q$-spectrum of the invariant measure of the effective
field in the dichotomous, symmetric 1D RFIM.

The detailed analysis of the
invariant measure of the effective field should be viewed as the preparation
of the study of the multifractal properties of the measure of the local
magnetisation in the RFIM which essentially is a convolution of the invariant
measure of the effective field with a distorted version of itself. The
transitions in the $D_q$-spectrum of the invariant measure of the effective
field have direct counterparts in the $D_q$-spectrum of the measure of the
local magnetisation and are thus gaining a direct physical significance. We will
address this subject in forthcoming work.

\ack Two of us (H.\;P. and U.\;B.) acknowledge discussions with Dr.\;Adrian
Lange in a very early stage of this work. The work was partially supported by
the Gra\-du\-ier\-ten\-kol\-leg ``Quantenfeldtheorie: Mathematische Struktur
und Anwendungen in der Elementarteilchen- und Festk\"orperphysik'' (DFG).
Thanks are due to anonymous referees for valuable comments.

\begin{appendix}
\section{Relations between scaling properties}
\subsection{Relation between the scaling at $\xsgn$ and at $\fmsgn(\xsgn)$}
\label{appa}
In this Appendix we show that the assumed scaling relation
\begin{align}
 \lim_{\varepsilon \to 0} \frac{P_\infty(\xsgn +
  \frac{\varepsilon}{2}) -
  P_\infty(\xsgn -
  \frac{\varepsilon}{2})}{\varepsilon^{\asgn+1}} =: k 
\end{align}
implies that the limit
\begin{align}
\lim_{\varepsilon \to 0} \frac{P_\infty\big(\fmsgn
  (\xsgn + \frac{\varepsilon}{2})\big) -
  P\big(\fmsgn(\xsgn -
  \frac{\varepsilon}{2})\big)}
{\big((\fmsgn)' (\xsgn) \, \varepsilon
  \big)^{\asgn +1}} \label{appe1}
\end{align}
exists and is equal to $k$. In the following we drop the indices
$\{\sigma\}_n$ and $\infty$ to improve readability and denote the expression 
in (\ref{appe1}) as $Q(\varepsilon)$. Applying
the mean value theorem to $f^{-1}$ in $Q(\varepsilon)$ and using $f^{-1}
(x^*) = x^*$ yields
\begin{align}
Q(\varepsilon) 
= \frac{P\big(x^* + (f^{-1})'
  (x^* + \delta_1) \frac{\varepsilon}{2}\big) - P\big(x^* - (f^{-1})'
  (x^* - \delta_2) \frac{\varepsilon}{2}\big)}{\big( (
  f^{-1})' (x^*) \, \varepsilon
  \big)^{\alpha +1}}   
\end{align}
with some $\delta_1, \delta_2 \in [0, \frac{\varepsilon}{2}]$.
Now let $(f^{-1})'_{\text{min}}$ be the minimum of $(f^{-1})' (x^* +
\delta_1)$ and $(f^{-1})' (x^* -
\delta_2)$. Because $f^{-1}$ is strictly monotonously growing we have
$(f^{-1})'_{\text{min}} > 0$.
Using the fact that $P$ is monotonously growing as well we get the
  lower estimate
\begin{align}
Q(\varepsilon) &\geq \frac{P\big(x^* + (f^{-1})'_{\text{min}}
  \frac{\varepsilon}{2}\big) - P\big(x^* - (f^{-1})'_{\text{min}}
  \frac{\varepsilon}{2}\big)}
{\big((f^{-1})' (x^*) \, \varepsilon \big)^{\alpha +1}} \\
&= \frac{P\big(x^* + (f^{-1})'_{\text{min}}
  \frac{\varepsilon}{2}\big) - P\big(x^* - (f^{-1})'_{\text{min}}
  \frac{\varepsilon}{2}\big)}
{((f^{-1})'_{\text{min}} \varepsilon)^{\alpha +1}} 
\cdot \bigg(
\frac{(f^{-1})'_{\text{min}} \varepsilon}{(f^{-1})'
  (x^*) \, \varepsilon} \bigg)^{\alpha + 1} .
\end{align}
The quotient of $(f^{-1})'_{\text{min}}$ and $(f^{-1})'
  (x^*)$ converges to $1$ and thus 
\begin{align}
\lim_{\varepsilon \to 0} Q(\varepsilon) \geq \lim_{\varepsilon \to 0} \frac{P\big(x^* + (f^{-1})'_{\text{min}}
  \frac{\varepsilon}{2}\big) - P\big(x^* - (f^{-1})'_{\text{min}}
  \frac{\varepsilon}{2}\big)}
{((f^{-1})'_{\text{min}} \varepsilon)^{\alpha +1}} = k .
\end{align}
Using the maximum of $(f^{-1})' (x^* +
\delta_1)$ and $(f^{-1})' (x^* -
\delta_2)$ instead of the minimum we get in the same fashion the upper estimate 
\begin{align}
\lim_{\varepsilon \to \infty} Q(\varepsilon) \leq k.
\end{align}
Both estimates together give the conjectured result $\lim_{\varepsilon
  \to 0} Q(\varepsilon) = k$.

\subsection{Relation between $\alpha(x)$ and $\alpha(f_\sigma^{-1}(x))$}
\label{appb}
The Frobenius-Perron equation for the invariant distribution $P_\infty$
induces the equality $\alpha(x)=\alpha(f_\sigma^{-1}(x))$ for any $x \in
\text{supp} \, \mu$ which is not in the overlap $O$.

The proof is straightforward. Let $x \in \text{supp} \,\mu$ be a point which
is not in $O$ and $\sigma \in \{-,+\}$ an arbitrary sign.  Then, dropping
again the index $\infty$,
\begin{align}
  \alpha(x)&= \lim_{\varepsilon \to 0} \frac{\ln(P(x+\frac{\varepsilon}{2})- P(x-
      \frac{\varepsilon}{2}))}{\ln \varepsilon} \\
    &= \lim_{\varepsilon \to
      0}\frac{\ln(\frac{1}{2}P(f_\sigma^{-1}(x+\frac{\varepsilon}{2}))-
        \frac{1}{2}P(f_\sigma^{-1}(x- \frac{\varepsilon}{2})))}{\ln \varepsilon}
\end{align}
because of the Frobenius-Perron equation. We now use the mean value theorem
for $f_\sigma^{-1}(x+\frac{\varepsilon}{2})$ and
$f_\sigma^{-1}(x-\frac{\varepsilon}{2})$ to obtain
\begin{align}
  &= \lim_{\varepsilon \to 0} \frac{-\ln 2}{\ln \varepsilon} +
  \frac{\ln(P(f_\sigma^{-1}(x)+ (f_\sigma^{-1})'(x+\delta_1)
  \frac{\varepsilon}{2}) -
  P(f_\sigma^{-1}(x)-(f_\sigma^{-1})'(x-\delta_2)\frac{\varepsilon}{2}))}{\ln
  \varepsilon} 
\end{align}
with some $\delta_1, \delta_2 \in [0, \frac{\varepsilon}{2}]$. Using again the
notation $(f_\sigma^{-1})'_{\text{min}}$ for the the minimum of
$(f_\sigma^{-1})' (x + \delta_1)$ and $(f_\sigma^{-1})' (x - \delta_2)$
and defining $\varepsilon ' := (f_\sigma^{-1})'_{\text{min}} \cdot \varepsilon$ we get
the inequality
\begin{align}
  \alpha(x) \geq \lim_{\varepsilon' \to 0} \frac{\ln(P(f_\sigma^{-1}(x) +
  \frac{\varepsilon '}{2}) - P(f_\sigma^{-1}(x) -
  \frac{\varepsilon'}{2}))}{-\ln(f_\sigma^{-1})'_{\text{min}} + \ln \varepsilon'}
&= \alpha(f_\sigma^{-1}(x)) .
\end{align}
Using the maximum $(f_\sigma^{-1})'_{\text{max}}$ of the derivatives instead of the
minimum we get by the same token $\alpha(x) \leq
\alpha(f_\sigma^{-1}(x))$. Therefore equality follows. 

\section{Independence of (\ref{indepsum}) from the choice of \boldmath$ x_i^{\sst
    (n)}$}
\label{app2}
In this Appendix we show that
\begin{align}
  \lim_{n \to \infty} \frac{1}{n}\Big(\sum_{i=1}^n \ln \phi(x_i^{\sst (n)}) -
  \sum_{i=1}^n \ln \phi(\tilde{x}_i^{\sst (n)})\Big) = 0 \label{averag}
\end{align}
for any strictly positive differentiable function $\phi$, any
symbolic sequence $\{\sigma\}$ and any choice of $x_i^{\sst (n)},
\tilde{x}_i^{\sst (n)} \in I_{\{\sigma\}_i}$, $i \in \{1, \ldots, n\}$, $n
\in \N$. Let $\phi$, $\{\sigma\}$, $x_i^{\sst (n)}$ and $\tilde{x}_i^{\sst (n)}$ be
given and set $\varepsilon_n := |I_{\{\sigma\}_n}|$. Then, using the mean
value theorem,
\begin{align}
  \phi(\tilde{x}_i^{\sst (n)}) = \phi(x_i^{\sst (n)}) + \phi'(x_i^{\sst (n)}
  +\delta_i^{\sst (n)})
  (\tilde{x}_i^{\sst (n)} - x_i^{\sst (n)} )
\end{align}
with some $\delta_i^{\sst (n)}, |\delta_i^{\sst (n)}| < \varepsilon_i$. Thus
\begin{align}
  \ln \phi(\tilde{x}_i^{\sst (n)}) - \ln \phi(x_i^{\sst (n)}) = \ln \Big( 1 +
  \frac{\phi'(x_i^{\sst (n)} + \delta_i^{\sst (n)})}{\phi(x_i^{\sst (n)})}
  \big(\tilde{x}_i^{\sst (n)} - x_i^{\sst (n)}\big)\Big) .
\end{align}
We denote the finite constant $\max \big\{ \frac{\phi'(x)}{\phi(x)} : x \in
I\big\}$ by $Q_{\text{max}}$. For sufficiently large $i \in \N$ the
expression $1- Q_{\text{max}}\, |\tilde{x}_i^{\sst (n)} - x_i^{\sst (n)}|$ is
positive and we get
\begin{align}
  & \ln \big(1- Q_{\text{max}}\, |\tilde{x}_i^{\sst (n)} - x_i^{\sst
    (n)}|\big) \; \leq \; \ln \phi(\tilde{x}_i^{\sst (n)}) - \ln \phi(x_i^{\sst
    (n)}) \; \leq \; \ln \big(1+ Q_{\text{max}}\, |\tilde{x}_i^{\sst (n)} -
  x_i^{\sst (n)}| \big) .
\end{align}
With $|\tilde{x}_i^{\sst (n)} - x_i^{\sst (n)}| \leq
\varepsilon_i$ this yields
\begin{align}
  \ln (1 - Q_{\text{max}}\, \varepsilon_i) \leq \ln \phi(\tilde{x}_i^{\sst (n)}) -
  \ln \phi(x_i^{\sst (n)}) \leq  \ln (1 + Q_{\text{max}}\, \varepsilon_i)
\end{align}
implying that the difference $\ln \phi(\tilde{x}_i^{\sst (n)}) -
\ln \phi(x_i^{\sst (n)})$ converges to zero. A standard argument then shows that
the average also converges to zero, i.e. that (\ref{averag}) is true.
 With $\phi= A'$ this yields the alleged independence of
(\ref{indepsum}) from the choice of $x_i^{\sst (n)}$.

\end{appendix}

\end{document}